\documentclass{aa}

\usepackage{graphicx}
\usepackage{txfonts}
\usepackage{bm}
\usepackage{natbib}
\bibpunct{(}{)}{;}{a}{}{,}
\begin{document}

   \title{The meso-structured magnetic atmosphere}

   \subtitle{A stochastic polarized radiative transfer approach}

   \author{T.A. Carroll
          \and
          M. Kopf
          }

   \offprints{T.A. Carroll}

   \institute{Astrophysikalisches Institut Potsdam,
              An der Sternwarte 16, D-14482 Potsdam\\
              \email{tcarroll@aip.de}
             }

   \date{Received December 19, 2006; accepted March 16, 2007}

   \abstract{
We present a general radiative transfer model which allows the Zeeman diagnostics 
of complex and unresolved solar magnetic fields.
Present modeling techniques still rely to a large extent on a-priori assumptions about the
geometry of the underlying magnetic field. 
In an effort to obtain a more flexible and unbiased approach we pursue a rigorous 
statistical description of the underlying atmosphere. 
Based on a Markov random field model the atmospheric structures are
characterized in terms of probability densities and spatial correlations.
This approach allows us to derive a stochastic transport equation for polarized light 
valid in a regime with an arbitrary fluctuating magnetic field on finite scales. 
One of the key ingredients of the derived stochastic transfer equation is the 
correlation length which provides an additional degree of freedom 
to the transport equation and can be used as a diagnostic parameter to 
estimate the characteristic length scale of the underlying magnetic field. 

It is shown that the stochastic transfer equation represents a natural 
extension of the (polarized) line formation under the micro- and macroturbulent 
assumption and contains both approaches as limiting cases. 
In particular, we show how in an inhomogeneous atmosphere asymmetric 
Stokes profiles develop and that the correlation length directly controls the 
degree of asymmetry and net circular polarization (NCP). 
In a number of simple numerical model calculations we demonstrate the  
importance of a finite correlation length for the polarized
line formation and its impact on the resulting Stokes line profiles.

\keywords{Radiative Transfer -- Line: formation -- Line: profiles --
          Sun: photosphere -- Sun: magnetic fields
         }
   }

   \maketitle
%

\section{Introduction}

   The entire solar photosphere exhibits a rich structure of large and small scale
   magnetic features like sunspots, pores,  faculae or plages.
   But except for sunspots and pores these magnetic fields are not
   spatially resolved with present telescopes, although these fields 
   clearly manifest themselves in high resolution spectropolarimetric observations. 
   With the improvement of spectropolarimetric sensitivity and 
   spatial resolution over the last years it became clear that 
   these unresolved magnetic fields are much more ubiquitous than previously thought. 
   This raises the question of the significance of these
   elusive and complex magnetic fields for the solar magnetism in general
   \citep{Schri03,SA04} and how these magnetic fields can be appropriately 
   investigated by means of spectropolarimetric observations and Zeeman 
   diagnostics. 
      
   The concept of magnetic flux tubes as the building blocks of solar surface magnetism 
   has surely played a key role in our understanding of the unresolved magnetic field 
   of the solar photosphere \citep{Stenflo76}. 
   A wealth of diagnostic techniques based on the flux tube concept have been developed 
   over the years and greatly improved our physical 
   insight into the nature of the photospheric magnetism \citep[e.g.][]{Sol93b}. 
   The interpretation of the Zeeman induced spectral line polarization in the context of 
   the flux tube model 
   rests on the idea of tube-like magnetic structures surrounded by an effectively
   non-magnetic atmosphere. A so called 1.5-dimensional radiative
   transfer model is then applied where a number of rays (line-of-sights) piercing 
   through the 2- or 3-dimensional geometry of the model atmosphere in order to obtain its 
   spectral signature \citep{Stenflo94}. 
   From the standpoint of radiative transfer 
   this approach provides a rather macroscopic treatment of the problem,
   as each line-of-sight possesses its own individual atmospheric structure
   and the averaging process for all
   line-of-sights (LOS) is performed after the actual integration of the transfer
   equation.  
   Another problem here is, a relatively detailed knowledge about the number density 
   and the geometrical arrangement of the flux tubes
   is needed in advance.  
   
   A quite different approach to characterize the small scale and unresolved nature of 
   the photospheric magnetic field was put forward by \citet{SA96}.
   In an attempt to reconcile the ubiquity of asymmetric Stokes V profiles
   with the underlying magnetic field structure, they postulated 
   the microstructured magnetic atmosphere (MISMA).
   In that model the degree of fluctuation and intermittency is much higher than 
   in the classical flux tube picture and the magnetic field is assumed to be
   structured over scales much smaller than the mean free path of photons. 
   This approach could successfully reproduce many of the observed Stokes V profile
   asymmetries in quiet and active solar regions \citep{SA97,SA00}. 
   In terms of radiative transfer modeling the MISMA approximation is a statistical
   approach which requires that the photons undergo a rapid and 
   random fluctuation on smallest scales along their trajectory (line-of-sight).
   If the fluctuation rate per length scale is much higher than the (true) absorption rate
   this scenario represents the classical microturbulent approach and allows a simplified 
   treatment of the radiative transfer. 
   The statistical averaging process (convolution) over all atmospheric structures 
   can then be performed before the actual integration of the transfer equation. 
   Despite its appealing simplicity in the way this approach treats the
   radiative transfer the idealized assumptions about the underlying atmosphere 
   strongly limits the application of this approach.
   As soon as only one of the ensemble structures exceeds the micro-scale criteria
   (and we will see later that this happens at surprisingly small scales) 
   the microturbulent approximation fails. Moreover, magnetic microstructures can not in general 
   be describe in the microturbulent limit. As elongated thin magnetic structures
   close to equipartition have a preferential vertical alignment due to buoyancy forces
   the path length of an individual line-of-sight that traverse through such a 
   magnetic structure depends on the angle between the line-of-sight and the axis of the 
   structure. The same structures that manifest themselves as microstructures in
   observations near the solar limb can become macrostructures in 
   disk-center observations. This simply reflects the fact that the
   vertical and horizontal dimension of flux structures in the solar photosphere 
   are very different.
   
   Over the last decade numerical magneto-convection 
   simulations have reached a level of realism where they can provide 
   valuable information about the three-dimensional structure of the photospheric
   plasma \citep{Stein98,Steiner98,Sch03,Vog03,Vog05,Schaf05,Stein06}. 
   Recent magneto-convection simulations of plage like regions have shown that 
   flux expulsion and convective field amplification may lead to structures in the
   supergranular network which resemble thin magnetic flux tubes or 
   flux sheets \citep{Vog05}. For the quiet sun, numerical simulations
   indicate a lesser degree of organization with more fragmented and 
   incoherent structures \citep{Schaf05,Stein06}.
   These simulations suggests that neither a predefined static macrostructured nor a pure
   microturbulent approach is an appropriate starting point for the interpretation of 
   spectropolarimetric observations.
   The magnetic field seems to comprise a broad range of structures which span
   from micro- (over meso-) to macro-scales. 
   This is picture is supported by recent Zeeman and Hanle based diagnostics
   which are consistent with a continuous magnetic field strength and flux distribution 
   \citep{Kho03,Dom03b,Bueno04}.

   So the present paper is an attempt to enhance the diagnostic capabilities 
   of the Zeeman effect and to 
   bridge the gap between the macro- and microstructured paradigm to obtain 
   a more general mesostructured approach.
   Our approach is based on a statistical description of the underlying atmosphere and 
   its relevant parameters in terms of a Markov random field. One of the aims is to
   explicitly incorporate the spatial coherences and to account 
   for the finite character of the underlying magnetic field. 
   Although the model atmosphere is inherently non-plane parallel this approach 
   allows us to formulate the transfer of polarized light with the
   help of a one dimensional stochastic transport equation.
   
   The idea of a stochastic radiative transfer approach is by no means new, 
   several attempts were made to describe the line formation in random velocity fields,
   for example \citet{Auv73,Gai74,Fri76,Mag85,Gu95,Nik97}
   The literature for polarized line formation in stochastic
   magnetic media is considerably shorter, there are only a few attempts known to the
   authors by \citet{Domke79,Lan04} and very recently by \citet{Car03,Car05a,Car05b,Car06} 
   and \citet{Fri05,Fri06}. The latter authors nicely demonstrated how the finite
   character of the underlying structures affect the line formation.
   But unlike most of the aforementioned authors we will not consider the solution 
   under some limiting aspects, our intention is rather to derive
   a general statistical description of the magnetized atmosphere and then, using this 
   description, to derive a (stochastic) transfer equation for polarized light.
   
   This paper is organized as follows: In Sect. \ref{Mesma} we introduce the general
   statistical concept of the mesostructured magnetic atmosphere. A model is presented 
   which explicitly takes into account the spatial coherency of the atmospheric parameters
   by means of a Markov random field.
   Based on that description, we derive in Sect. \ref{StochTrans} the stochastic transport 
   equation for polarized light. We present a formal solution in terms of a modified
   evolution operator and show that the microturbulent and the macroturbulent
   approximations are special limits of our more general mesostructured approach. 
   In Sect. \ref{StochTrans} we also show that the origin of asymmetric 
   Stokes profiles and a net-circular-polarization (NCP) are natural
   consequences in a mesostructured magnetic atmosphere and that the degree of asymmetry
   depends directly on the characteristic length scale (correlation length) of the 
   underlying magnetic structures.
   In Sect. \ref{Experiments} we give a brief description of the numerical realization of
   our stochastic approach. Then a number of numerical experiments follow,
   where we demonstrate how the arrangement of the underlying atmosphere and in particular 
   the correlation length of the magnetic structures have a decisive impact on the line 
   formation process and on the appearance of the resulting Stokes profiles.
   Sect. \ref{Summary} finally gives a brief summary and draws the main conclusions. 


\section{The Mesostructured Magnetic Atmosphere}
\label{Mesma}
Due to solar convection we know that the photospheric
plasma has a rather complicated structure in terms of its temperature and velocity
distribution. This dynamic behavior also influences the magnetic field, in
particular in a regime where the plasma $\beta$ is close to one.
It is this complexity why we have chosen to pursue a rigorous statistical approach 
to describe the atmosphere and the underlying magnetic field.

In the following we will present a discrete description of the atmosphere, following 
the concept of numerical magnetoconvective simulations, we approximate the atmospheric structure 
by a discrete 3-dimensional lattice structure with a regular arrangement of 
nodes. To each node we will assign a number of physical parameters (a state vector)
which describe the physical conditions at the particular locations of the nodes.
Each node is moreover associated with a multivariate probability density function 
and a neighborhood relation. 
This neighborhood relation will be given in terms of a Markov model 
which then allows us to derive, for an arbitrary trajectory or line-of-sight
a differential equation which describes the 
evolution of the probability density function.
This differential equation forms the basis for Sect. \ref{StochTrans}, where we
describe the line formation as a stochastic process and derive the 
stochastic transport equation for polarized light.

\subsection{The atmosphere as a Markov random field}
\label{MarkovField}

Let us begin by introducing the state vector $\bm{B}$  for an arbitrary position (node)
in the atmosphere as a vector quantity which includes all relevant atmospheric 
parameters, 
\begin{equation}
\bm{B}^* = (\alpha_{1}^*, \alpha_{2}^*, \dots \alpha_{n}^*)^T \: . \label{struct_vec}
\end{equation}
The vector $\bm{B}$ shall contain atmospheric parameters like temperature, velocity, magnetic field etc..
We assume, that the parameters $\alpha_{n}$ are random variables, hence the vector $\bm{B}$ 
represents a multivariate random variable or a random vector respectively, for which we define
a probability density function (pdf) of the form
\begin{equation}
p(\bm{B}) = p(\alpha_{1},\alpha_{2}, \dots \alpha_{n}) \: ,
\label{basic_pdf}
\end{equation}
where the probability $P(\bm{B} \leq \bm{\hat{B}} \leq \bm{B}+ d\bm{B})$ of finding a realization of 
$\bm{\hat{B}}$ within $\bm{B} \leq \bm{\hat{B}} \leq \bm{B}+ d\bm{B}$ is
\begin{eqnarray}
P(\bm{B} \leq \bm{\hat{B}} \leq \bm{B}+ d\bm{\hat{B}}) = p(\alpha_1,\dots,\alpha_n) \;  d\alpha_1 \dots d\alpha_n \nonumber \\
= p(\bm{B}) \; d\bm{B} \: , \hspace{2.15cm}
\end{eqnarray} 
and which satisfy the following normalization condition: 
\begin{equation}
p(\bm{B}) \geq 0 \: \: \mbox{and} \: \int_{-\infty}^{\infty} \; p(\bm{B}) \; d\bm{B} \; =  \; 1 \; .
\label{normcond}
\end{equation} 
Here, we have assumed that $\bm{B}$ is a real-valued random vector and we want to keep this
notation in the following without loss of generality. 
For the sake of brevity and clarity, we also will not distinguish between the
stochastic variable and its realizations (states) hereafter, furthermore the integration 
limits will be omitted as long as the integration is performed over the entire domain
or state space respectively. 

As mentioned in the introduction, we will now consider a spatially discretized 
approximation of the atmosphere in terms of a three dimensional uniform grid structure. 
The grid, hereafter called the random field, is made up by a number of 
grid points (nodes) where each node is associated with a random atmospheric vector 
$\bm{B}_i$, and its respective probability density function  
\begin{equation}
p_{s_i}(\bm{B}_i) = p(\bm{B}_i,s_i) = p(\alpha_{i_1},\alpha_{i_2}, \dots \alpha_{i_n},s_i) \: .\label{atmos_pro_pdf}
\end{equation}
The index $s_i$ specifies the spatial position of the node.
In order to obtain a general probabilistic description of the entire 
field, we need to take into account that the 
random vectors $\bm{B}_i$ at their specific positions $s_i$ might be correlated among each 
other. The statistical state of the grid is 
therefore described in its most general form by a joint probability density for the entire field
$p(\bm{B}_{1},s_1;\dots;\bm{B}_{n},s_n)$. This joint pdf or n-point distribution function can 
also be expressed in terms of the respective conditional probability density function 
\begin{eqnarray}
p(\bm{B}_{1},s_1;\dots;\bm{B}_{n},s_n) = p(\bm{B}_{1},s_1 \mid \bm{B}_{2},s_2; \dots ;\bm{B}_{n},s_n)
 \: \hspace{0.6cm} \nonumber \\
\times \: p(\bm{B}_{2},s_2;\dots;\bm{B}_{n},s_n) \; , 
\label{atmos_cpdf}
\end{eqnarray}  
which emphasizes the spatial dependency (correlation) of the field. 
The state vector $\bm{B}_{1}$ at $s_1$ will in general depend on other state vectors 
within a certain region or radius around  $s_1$. 
In general it will be difficult to formulate an approximation for 
the joint probability density function or conditional probability density 
(\ref{atmos_cpdf}) in terms of an appropriate n-point correlation function. 
To reduce the complexity of the random field the
simplest assumption is that of a completely uncorrelated field. 
Thus, the realization of a state vector $\bm{B}_{i}$ at the position $s_i$ 
is completely independent and uncorrelated from all other realizations at other positions. 
With this assumption the joint probability density (\ref{atmos_cpdf}) 
factorizes into the products of the individual one-point probability densities of the 
respective state vectors, 
\begin{eqnarray}
p(\bm{B}_{1},s_1;\dots;\bm{B}_{n},s_n) = p(\bm{B}_{1},s_1)p(\bm{B}_{2},s_2) \dots p(\bm{B}_{n},s_n) \nonumber \\
= \prod_i^n p(\bm{B}_{i},s_i) \; . \hspace{2.45cm}
\label{atmos_uncorrel}
\end{eqnarray} 
This description of the atmosphere in terms of a spatially uncorrelated field where 
the atmospheric state vectors $\bm{B}_i$ are exclusively described by their respective 
one-point pdf, would immediately lead us to a microturbulent approach. 
Our intention here is, however, to go beyond this microturbulent concept in order 
to take -- at least to the lowest order -- 
correlation effects into account. In the following we will therefore 
consider the particular type of a Markovian field structure. 
A  Markov random field is characterized by a rather simple neighborhood relation 
which describes the spatial correlation of each node in the field with its 
neighboring nodes.  
The neighborhood $N_i$ at the position $s_i$ can be defined as follows
\begin{equation}
N_i \equiv \left \{ s_j \: : \: \parallel s_i - s_j \parallel \leq r \right \} \; ,
\label{neighborhood}
\end{equation} 
where $r$ gives the spatial distance or radius between $s_i$ and the neighboring 
points $s_j$. The Markov property of the random field can then be defined such that
for each state vector $\bm{B}_i$ at $s_i$ the following 
conditional probability density function holds
\begin{equation}
p(\bm{B}_{i},s_i \mid \bm{B}_{j},s_j)
\; ; \mbox{where} \; i \neq j \; \mbox{and} \; s_j \in N_i  \; .\label{atmos_markov_cpdf}
\end{equation}
The conditional probability of the state vector $\bm{B}_i$ is thus only correlated to
its neighboring states. 
The underlying assumption here is that all higher order spatial correlations are 
sufficiently small (and rapidly decay to zero) compared to first order correlation effects
such that the atmospheric vector $\bm{B}_i$ at $s_i$  decouples from situations 
(states) which are further away than the immediate neighborhood $N_i$.

\subsection{A sample path through the Markov field}
\label{MarkovProcess}

In this work we will concentrate on the radiative transfer in a non-scattering 
atmosphere and therefore let us consider a particular trajectory -- a sample path -- 
through the above described random field. 
The sample path is described by a one-dimensional, unidirectional and contiguous 
sequence (path or trajectory) with 
$s_1 < s_2 < \dots < s_n$ through the random Markov field, where $s_i$ now determines 
the node position along our sample path.  
Note, that the state space vector $\bm{B}$ can either be discrete or continuous, 
without loss of generality and for 
mathematical convenience we assume in the following a continuous state space.
Along the trajectory we encounter a sequence of realizations with 
$\bm{B}_1,s_1;\bm{B}_2,s_2;...;\bm{B}_n,s_n$ where the random vector $\bm{B}_i$ in general
adopts different values at different positions $s_i$. 
This stochastic sequence along the trajectory defines a stochastic process
where the state vector $\bm{B}$ has a distinct dependence on the spatial position $s$.
From the Markov property of the random field we know that there exist
a conditional probability density for the vector $\bm{B}_i$ at each position $s_i$ in the atmosphere 
which relates $\bm{B}_i$ to the state vectors of the neighborhood. 
For the stochastic process along our sample path the three 
dimensional neighborhood at each position reduces to a one dimensional relation. 
In particular for our unidirectional path this means that the state $\bm{B}_i$ at $s_i$ 
is correlated only to its predecessor state $\bm{B}_{i-1}$ at $s_{i-1}$. 
This represents the typical property of a Markov process \citep{vanK92} 
and allows us to write for the conditional probability densities at $s_n$  
\begin{equation}
p(\bm{B}_n,s_n \mid \bm{B}_{n-1},s_{n-1}; \dots ;\bm{B}_1,s_1) \: = \:
p(\bm{B}_n,s_n \mid \bm{B}_{n-1},s_{n-1}), \label{markov_cpdf}
\end{equation}  
where $s_n$ and $s_{n-1}$ are neighboring points along our sample path. 
The conditional probability which, in the framework of Markov processes, 
is also called transition probability expresses the property that 
the probability for a transition from the state $\bm{B}_{n-1}$ at $s_{n-1}$
to a state  $\bm{B}_n$ at $s_n$ only depends on the condition (state) 
at $s_{n-1}$ and not on the (spatial) history prior to $\bm{B}_{n-1}$.
As can be shown, the following equation must hold for a Markov process \citep{vanK92},
\begin{eqnarray}
p(\bm{B}_3,s_3 \mid \bm{B}_1,s_1) \: = \int p(\bm{B}_3,s_3 \mid \bm{B}_2,s_2) \: \hspace{1.9cm} \nonumber \\
\times \: p(\bm{B}_2,s_2 \mid \bm{B}_1,s_1) \: d\bm{B}_2 \: , 
\label{chapkol}
\end{eqnarray}
where the integration is performed over the entire
state space. This is the Chapman-Kolmogorov equation which states that the 
transition probability between two states $\bm{B}_1$ and $\bm{B}_3$ with $s_1 < s_3$ corresponds 
to the product of the transition probability between the initial state $\bm{B}_1$ and 
some intermediate state $\bm{B}_2$ as well as the transition from this intermediate state 
to the final state $\bm{B}_3$, integrated over all possible intermediate states.

In the following, we will concentrate on a homogeneous stochastic process.
A homogeneous process is characterized by a transition probability 
$p(\bm{B}_2,s_2 \mid \bm{B}_1,s_1)$ that only depends 
on the spatial distance $\Delta s = s_2 - s_1$ between the two states and not on 
their particular spatial positions. 
We therefore omit the indices and consider, without loss of generality, $s$ as a 
continuous parameter.

Let us now define a transition probability which obeys the Chapman-Kolmogorov
Eq. (\ref{chapkol}). To obtain such a model we have to make 
assumptions about the underlying (magnetic) structure of the atmosphere and how this
structure determines the short scale behavior of the transition probability.
Following the idea of thin magnetic flux tubes we can assume that the photosphere is
characterized by sharp transitions between individual magnetic structures 
and their non-magnetic surrounding. 
This picture may be justified by the extreme small boundary layers of 
thin magnetic flux tubes which are well below the photon mean free path \citep{Sch86}. 
The Markov process must therefore exhibit a distinct discontinuous character
where abrupt jumps between different atmospheric regimes along a trajectory occur.
A suitable definition of the transition probability for small $\Delta s$, between two 
states $\bm{B}'$ and $\bm{B}''$ can be obtained by the following discontinuous process
\begin{eqnarray}
p(\bm{B}'',s+\Delta s \mid \bm{B}',s) = e^{- \gamma(\bm{B}') \Delta s}
\; \delta(\bm{B}' -\bm{B}'') \; \hspace{1.5cm} \nonumber \\
+ \; (1 - e^{- \gamma(\bm{B}') \Delta s}) \: q(\bm{B}'') 
\: , \label{kubo}
\end{eqnarray}
where $\gamma(\bm{B}')$ is the fluctuation rate of the atmospheric structure $\bm{B}'$ and
$q$ a prescibed spatial independent probability density function. This probability 
density, which has to satisfy the normalization condition (\ref{normcond}), allows a prior characterization of the 
field and transition properties. Note, that the prescribed probability density $q$
is in general not the stationary (spatially-invariant) probability density of the process \citep{Briss74}.
The fluctuation rate can be expressed with the help of the correlation length $\lambda$ to give
\begin{eqnarray}
\gamma(\bm{B}) = \lambda^{-1}(\bm{B}) \: .
\label{corrlength}
\end{eqnarray}
As can be seen from Eq. (\ref{kubo}) the correlation length can be considered as 
the mean length scale of the structures.
The process (\ref{kubo}), which is sometimes called Kubo-Anderson process 
\citep{Fri76} describes how the probability of the state vector $\bm{B}''$  
depends on the spatial distance $\Delta s$ between $\bm{B}''$ and the initially
known vector $\bm{B}'$. 
The probability for staying in the initial state or atmospheric regime $\bm{B}'$ 
decreases exponentially with the distance $\Delta s$ while the probability for a  
transition to another atmospheric regime $\bm{B}''$  increases with 
$\Delta s$ and is weighted by the given probability density $q(\bm{B}'')$. 
As the conditional probability density of the process depends on the spatial
distance $\Delta s$ we can use the Kubo-Anderson process (\ref{kubo}) and 
the Chapman-Kolmogorov Eq. (\ref{chapkol}) to derive
the following integro-differential equation (see Appendix \ref{Appendix_Master})
which describes the evolution of the probability density for $\bm{B}$ along a 
particular trajectory.
\begin{eqnarray}
\frac{\partial p(\bm{B},s)}{\partial s}  \: = \:
\int w(\bm{B},\bm{B}',s) \: p(\bm{B}',s) \: d\bm{B}' \: \hspace{1.9cm} \nonumber \\
- \: \int w(\bm{B}',\bm{B},s) \: p(\bm{B},s) \: d\bm{B}' \: , 
\label{master_equation}
\end{eqnarray}
where $w$ is the transition rate (transitions per unit length scale) and is defined by (\ref{transrate}). 
This equation which is also called the master equation of the process describes
how the probability density evolves along a given trajectory through the atmosphere.
The first term on the r.h.s. represents a source or gain term for the probability,
which describes transitions from an arbitrary atmospheric state $\bm{B}'$ to the 
state $\bm{B}$ under consideration. The second term on the r.h.s. 
is a sink or loss term for the probability density and describes 
transitions from the state $\bm{B}$ to other available atmospheric states.
The master equation forms the basis for the following derivation of the 
stochastic transfer equation for polarized light.

\section{The stochastic transport equation for polarized light}
\label{StochTrans}
\subsection{Derivation of the stochastic transport equation}
\label{DerivStochTrans}

We now consider a particular photon trajectory along the line-of-sight
in a non-scattering atmosphere, which will allow us to describe the 
radiative transfer in terms of a stochastic process.
We begin by writing down the (deterministic) transport equation for polarized 
light along a given ray path coordinate $s$ in a plane parallel atmosphere for a 
frequency $\nu$, which is given by 
\begin{equation}
\frac{d\bm{I}_{\nu}}{ds} \: = \: \bm{K}_{\nu} \bm{I}_{\nu} \: + \: \bm{j}_{\nu} \;,
\label{d_transport}
\end{equation}
where $\bm{I}$ is the 4-dimensional Stokes vector $\bm{I} = {I,Q,U,V}^T$ which describes
the intensity and polarization state of the light \citep[e.g.][]{Stenflo94}.
For clarity, we will omit the frequency indices in the following.
The absorption matrix $\bm{K}$ and the emission vector $\bm{j}$ are given by
\begin{displaymath}
\bm{K} \: = \: \left (
\begin{array}{cccc}
\eta_{I} & \eta_{Q} & \eta_{U} & \eta_{V} \\
\eta_{Q} & \eta_{I} & \rho_{V} & -\rho_{U} \\
\eta_{U} & -\eta_{V} & \eta_{I} & \rho_{Q} \\
\eta_{V} & \rho_{U} & -\rho_{Q} & \eta_{I}
\end{array}
\right ) 
\label{absmatrix}
\end{displaymath}
and
\begin{equation}
\bm{j} \: = \: (j_I,j_Q,j_U,j_V)^T  = (\eta_{I},\eta_{Q},\eta_{U},\eta_{V})^T B\: ,
\label{emissvec}
\end{equation}
where T denotes the transpose of the emission vector and $B$ stands 
for the Planck function. Here we have assumed that the emission 
originates under local thermodynamic equilibrium (LTE). 
 For the standard case of a spectral line formed in
the presence of a magnetic field, the detailed expression for the entries of 
the absorption matrix $\bm{K}$ as well as for the emission vector $\bm{j}$ can 
be found for instance in \citet{Rees87}.

As our model atmosphere is described in statistical terms the 
observable quantity of interest in our case is now given by the 
mean value of the Stokes vector.
We therefore define the mean Stokes vector by the first moment of the 
probability density function $p(\bm{I},\bm{B},s)$. Two things are noteworthy here,
first, we have to include the spatial position $s$ into the
density function since the value of the density function in general depends on 
the spatial position in the atmosphere since both $\bm{I}$ and $\bm{B}$ depend on $s$.
Second, we have to include the Stokes vector itself into the statistical 
description, because the Stokes vector is coupled to the atmospheric 
conditions via the transport Eq. (\ref{d_transport}) and 
therefore the Stokes vector itself becomes a stochastic variable which 
follows the same stochastic process as the atmospheric vector $\bm{B}$. 
The mean Stokes vector at the position $s$ is then given by
\begin{equation}
<\bm{I}(s)> \: = \: \int \int \: \bm{I} \: p(\bm{I},\bm{B},s) \: d\bm{B} \: d\bm{I}.
\label{meanI1}
\end{equation}
We then separate the joint pdf $p(\bm{I},\bm{B},s)$ by using the conditional 
pdf $p(\bm{I}\mid\bm{B},s)$ to write Eq. (\ref{meanI1}) as
\begin{equation}
<\bm{I}(s)> \: = \: \int \int \: \bm{I} \: p(\bm{I} \mid \bm{B},s) \: p(\bm{B},s) \: d\bm{B} \: d\bm{I} \: .
\label{meanI2}
\end{equation}
The joint pdf can therefore be expressed as the product of the conditional pdf of the 
Stokes vector $\bm{I}$ which is conditioned on the atmospheric regime $\bm{B}$ at $s$ and the structural pdf 
of the atmospheric vector $\bm{B}$.
We can evaluate the conditional pdf $p(\bm{I}\mid\bm{B},s)$
by realizing that this conditional pdf represents the deterministic solution of the transport 
Eq. (\ref{d_transport}) under the specific atmospheric conditions $\bm{B}$. 
Thus expressing the conditional pdf with the help of the Dirac delta function we obtain
\begin{equation}
p(\bm{I} \mid \bm{B},s)\: = \: \delta(\bm{I} - \bm{I}(\bm{B},s)) \: = \: \delta(\bm{I} - \bm{\hat{I}}(s)) \; ,
\label{deltaI}
\end{equation}
where $\bm{\hat{I}}(s)$ is the formal solution of the initial value problem of Eq. (\ref{d_transport}) 
given the atmospheric conditions $\bm{B}$ at $s$. The Dirac delta function of the
Stokes vector has the following definition
\begin{eqnarray}
\delta(\bm{I} - \bm{\hat{I}}(s))  \hspace{6.5cm} \nonumber \\
= \delta(I - \hat{I}(s)) \; 
\delta(Q - \hat{Q}(s)) \; 
\delta(U - \hat{U}(s)) \; 
\delta(V - \hat{V}(s)) \: . 
\label{defdelta}
\end{eqnarray}
Eq. (\ref{meanI2}) can then be written as
\begin{equation}
<\bm{I}(s)> \: = \: \int \int \: \bm{I} \: \delta(\bm{I} - \bm{\hat{I}}(s)) \: p(\bm{B},s) \:  d\bm{B} \: d\bm{I} \: ,
\label{meanI3}
\end{equation}
Taking the derivative of (\ref{meanI2}) with respect to the 
path variable $s$ such that
\begin{equation}
\frac{\partial <\bm{I}>}{\partial s} \: = \frac{\partial}{\partial s} \: \int \int \bm{I} \: 
\delta(\bm{I} - \bm{\hat{I}}(s)) \: p(\bm{B},s) \: d\bm{B} \: d\bm{I} \: ,
\label{meanderiv1}
\end{equation}
and interchanging the differentiation and integration operators we get
\begin{eqnarray}
\frac{\partial <\bm{I}>}{\partial s} \: = \int \int \: \left [ \: \bm{I} \: \left ( \frac{\partial}{\partial s} \right.
\delta(\bm{I} - \bm{\hat{I}}(s)) \right ) \: p(\bm{B},s) \:  \hspace{1.2cm} \nonumber \\ 
+ \: \bm{I} \: \delta(\bm{I} - \bm{\hat{I}}(s)) \:
\left. \frac{\partial}{\partial s} \: p(\bm{B},s) \right ]\: d\bm{B} \: d\bm{I} \: .
\label{meanderiv2}
\end{eqnarray}
Taking further into account the differential form of the transport Eq. (\ref{d_transport}), 
we can write the derivative of the 
Dirac delta function (\ref{defdelta}) with respect to the path variable $s$ as
\begin{eqnarray}
\frac{\partial}{\partial s}  \delta(\bm{I} - \bm{\hat{I}}(s)) & = &
\frac{\partial \bm{\hat{I}}(s)}{\partial s} \; \nabla_{\bm{\hat{I}}} \; 
\delta(\bm{I} - \bm{\hat{I}}(s)) \nonumber \\ 
& = & (- \bm{K_B \hat{I}} + \bm{j_B}) \; \nabla_{\bm{\hat{I}}} \; \delta(\bm{I} -
\bm{\hat{I}}(s)) \nonumber \\
& = & - (- \bm{K_B \hat{I}} + \bm{j_B}) \; \nabla_{\bm{I}} \; 
\delta(\bm{I} - \bm{\hat{I}}(s)) \nonumber \\ 
& = & - \nabla_{\bm{I}} \left [ (- \bm{K_B I} + \bm{j_B}) \; \delta(\bm{I} - \bm{\hat{I}}(s)) 
\right ] \: ,
\label{derivdelta}
\end{eqnarray}
where $\bm{K_B}$ and $\bm{j_B}$ are the absorption matrix and emission vector 
valid in the regime $\bm{B}$, $\nabla_{\bm{I}}$ is the nabla operator for 
the Stokes vector which is defined as
\begin{equation}
\nabla_{\bm{I}} = \left ( 
\begin{array}{c} 
\frac{\partial}{\partial I} \\
\frac{\partial}{\partial Q} \\
\frac{\partial}{\partial U} \\
\frac{\partial}{\partial V} 
\end{array} \right ) \; .
\label{deltastokes}
\end{equation}
Relation (\ref{derivdelta}), allows us to write Eq. (\ref{meanderiv2}) as
\begin{eqnarray}
\frac{\partial <\bm{I}>}{\partial s} \hspace{6.9cm} \nonumber \\
= \int \int - \nabla_{\bm{I}} \left [ (- \bm{K_B I} + \bm{j_B}) \delta(\bm{I} - \bm{\hat{I}}(s)) \right ]
p(\bm{B},s) \:\bm{I} \: \: d\bm{I} \: d\bm{B}  \nonumber \\
+ \int \int \: \bm{I} \: \delta(\bm{I} - \bm{\hat{I}}(s)) \: \frac{\partial}{\partial s} 
p(\bm{B},s) \: \: d\bm{I} \: d\bm{B} . \hspace{0.0cm}
\label{meanderiv3}
\end{eqnarray}
If we now use the master Eq.(\ref{master_equation}) to express the spatial derivative of 
$p(\bm{B},s)$, Eq.(\ref{meanderiv3}) can be written as,
\begin{eqnarray}
\frac{\partial <\bm{I}>}{\partial s}  \hspace{6.9cm} \nonumber \\
\hspace{0.5cm} = \int \int - \nabla_{\bm{I}} \left [ (- \bm{K_B I} + \bm{j_B}) \: \delta(\bm{I} - \bm{\hat{I}}(s)) \right ] 
p(\bm{B},s) \bm{I} \: d\bm{I} \: d\bm{B}  \hspace{0.0cm} \nonumber \\ 
+ \int \int \left( \int w(\bm{B},\bm{B}') 
p(\bm{B}',s)  d\bm{B}' \right ) \bm{I} \: \delta(\bm{I} - \bm{\hat{I}}(s)) \: d\bm{I} \: d\bm{B} \hspace{0.0cm} \nonumber \\ 
- \int \int \left( 
\int w(\bm{B}',\bm{B}) p(\bm{B},s) d\bm{B}' \right ) \bm{I} \: \delta(\bm{I} - \bm{\hat{I}}(s)) \: d\bm{I} \: d\bm{B} .
\hspace{0.0cm}
\label{meanderiv4}
\end{eqnarray}
We then introduce the mean conditional Stokes vector (conditioned on the
atmospheric regime $\bm{B}$)
which is defined as
\begin{equation}
\bm{Y_B}(s) \: = \: \int \bm{I} \: p(\bm{I} \mid \bm{B},s) \: d\bm{I} \: .
\label{meancond1}
\end{equation}
The components of the mean conditional Stokes vector are given in analogy to the 
definition of the regular Stokes vector,
\begin{equation}
\bm{Y_{B}}= \left ( 
\begin{array}{c} 
Y_B^{I} \\
Y_B^{Q} \\
Y_B^{U} \\
Y_B^{V}
\end{array} \right ) \; .
\label{meancondpar}
\end{equation}
Taking into account Eq. (\ref{meanI2}) as well as the definition of
the mean conditional Stokes vector we can integrate the first term on the 
r.h.s. of Eq. (\ref{meanderiv4}) by parts to obtain the following equation
\begin{eqnarray}
\frac{\partial}{\partial s} \int \bm{Y_{B}} \: p(\bm{B},s) \: d\bm{B} \: = \: \int (- \bm{K_B Y_B} + \bm{j_B}) 
p(\bm{B},s) \: d\bm{B} \hspace{0.0cm} \nonumber \\ 
+ \int \left( \int w(\bm{B},\bm{B}') \: \bm{Y_{B'}} \: p(\bm{B}',s) \: d\bm{B}' \right ) d\bm{B} \hspace{0.0cm} \nonumber \\
- \int \left( \int w(\bm{B}',\bm{B}) \: \bm{Y_{B}} \: p(\bm{B},s)\: d\bm{B}' \right ) d\bm{B} \: , \hspace{0.0cm} 
\label{meancond2}
\end{eqnarray}
where we have made the reasonable assumption that the atmospheric vector $\bm{B}$ is independent from
the Stokes intensities. If we now neglect the outer integration
over the state space $\bm{B}$ and divide Eq. (\ref{meancond2})
by $p(\bm{B},s)$ as well as taking into account the definition of the transition rate (\ref{transrate}) 
we obtain the following integro-differential equation for the mean
conditional Stokes vector
\begin{eqnarray}
\frac{\partial \bm{Y_{B}}}{\partial s} \: = \: - \bm{K_B Y_{B}} + \bm{j_B} + 
\int \bar{\gamma}(\bm{B}') \bm{Y_{B'}} p(\bm{B}',s) \: d\bm{B}' \hspace{1.2cm} \nonumber \\ 
 - \int \gamma(\bm{B})  \bm{Y_{B}} q(\bm{B}') d\bm{B}' .
\label{meancond3}
\end{eqnarray}
Here, we have introduced the modified fluctuation rate $\bar{\gamma}(\bm{B})$ which is given by
\begin{equation}
\bar{\gamma}(\bm{B}') = \frac{q(\bm{B})}{p(\bm{B},s) \lambda(\bm{B}')} \: .
\label{modgamma}
\end{equation}
Eq. (\ref{meancond3}) is the transport equation for the mean conditional Stokes vector $\bm{Y_{B}}$
which can be solved by specifying the initial conditions at the bottom of
the atmosphere
\begin{eqnarray}
\bm{Y_{B_0}(s_0)} \: = \: \int \bm{I}_0 \: p(\bm{I}_0 \mid \bm{B}_0, s_0) \: d\bm{I}_0 \nonumber \hspace{1.0cm} \\
= \int \bm{I}_0 \: \delta(\bm{I}_0 - \bm{I}(s_0)) \:  d\bm{I}_0 \: ,\hspace{1.0cm}
\label{icond}
\end{eqnarray}
as well as an initial probability distribution $p(\bm{B}_0,s_0)$.
Eq. (\ref{meancond3}) can be simplified under the assumption that the 
fluctuation rate $\gamma$  does not depend on the particular magnetic regime
which then allows us to write
\begin{equation}
\frac{\partial \bm{Y_{B}}}{\partial s} \: = \: - \bm{K Y_{B}} + \bm{j}
+ \bar{\gamma} <\bm{I}(s)> - \gamma \bm{Y_{B}} \: .
\label{meancond4}
\end{equation}
The mean \emph{unconditional} Stokes vector at the top of the atmosphere $s_{t}$ -- our observable quantity -- can 
then be readily obtained by a final integration of the mean conditional Stokes vector and the prescribed probability density q
over the entire state space of $\bm{B}$,
\begin{equation}
<\bm{I}(s_{t})> \: = \: \int \bm{Y_B}(s_{t}) \: q(\bm{B}) \: d\bm{B} \: .
\label{meanstokes}
\end{equation}
Eq. (\ref{meancond3}) describes the evolution of the mean conditional Stokes vector 
through the atmosphere from which the mean Stokes vector, the macroscopic (observable) 
quantity can be obtained. The stochastic transport equation for the mean 
conditional Stokes vector (\ref{meancond3}) has the general form of a stochastic master 
equation \citep{vanK92} and describes the transport of photons which are subject 
to the familiar processes of absorption (thermalization) and emission as well as two 
processes which describes the flow (fluctuation) between the different atmospheric 
regimes or components of the atmosphere. To understand the two new terms in the
stochastic transport equation it is important to realize that
Eq. (\ref{meancond3}) is conditioned on one particular atmospheric regime $\bm{B}$ 
such that the integration process is entirely performed within that particular 
regime.
Hence ,the two new terms in the stochastic transfer equation statistically 
couples the conditional transport equation for $\bm{B}$ to the other atmospheric 
regimes. In detail, 
the third term on the r.h.s. of Eq. (\ref{meancond3}) gives the amount of intensity or 
photons which enters per unit length scale from all other atmospheric regimes into the
regime $\bm{B}$ under consideration, this term therefore describes an additional 
\emph{statistical} inflow or source term, while the fourth term describes the loss of 
intensity per unit length scale due to transitions of photons to other possible regimes 
and therefore adds an additional \emph{statistical} absorption. Due to the resemblance to
the transport equation in a scattering medium both 
statistical terms can be considered as statistical scattering terms
which describe the scattering of photons in and from the regime $\bm{B}$.  
The stochastic transfer Eq. (\ref{meancond3}) can be written in a more
compact form if we combine the statistical source term with the emission vector $\bm{j_B}$
and the statistical absorption with the absorption matrix $\bm{K_B}$.
We can then define a modified (conditional) absorption matrix $\bm{\tilde{K}_B}$ such as
\begin{equation}
\bm{\tilde{K}_B} \: = \: \bm{K_B} \: + \: \bm{1} \gamma \; , 
\label{modabsorb}
\end{equation}
where $\bm{1}$ is the unit $4 \times 4$ matrix. The modified (conditional) 
emission vector $\bm{\tilde{j}}$ can be defined as
\begin{equation}
\bm{\tilde{j}_B} \: = \: \bm{j_B} \: + \: \int \bar{\gamma}(\bm{B}) \: \bm{Y_{B'}}  \: p(\bm{B}',s) \: d\bm{B}' \: . 
\label{modemission}
\end{equation}
This allows us to write the stochastic transport Eq. (\ref{meancond3}) in the following form
\begin{eqnarray}
\frac{\partial \bm{Y_{B}}}{\partial s} \: = \: - \bm{\tilde{K}_B Y_{B}} 
+ \bm{\tilde{j}_B} \: .
\label{meancond5}
\end{eqnarray}
In the case of a discrete state space $\bm{B}$ we can
obtain the discrete form of the stochastic transport Eq. (\ref{meancond3}) from 
the discrete master equation Eq. (\ref{app_disc_master}) which gives 
\begin{eqnarray}
\frac{\partial \bm{Y_{B_m}}}{\partial s} \: = \: - \bm{K_{B_m} Y_{B_m}} + \bm{j_{B_m}} + 
\sum_n \bar{\gamma}(\bm{B_{n}}) \: \bm{Y_{B_n}} \: p(\bm{B_n},s)
\hspace{0.8cm} \nonumber \\
 - \sum_n \gamma(\bm{B_{m}}) \: \bm{Y_{B_m}} \: q(\bm{B_n})  \: .
\label{meancond6}
\end{eqnarray}

\subsection{The formal solution}
\label{Formal}
A formal solution for the mean conditional Stokes vector can be obtained in analogy to 
the formal solution of the polarized radiative transfer equation \citep{Lan85}.
Taking into account the definition of the modified absorption matrix (\ref{modabsorb}) 
the transfer equation (\ref{meancond5}) in a purely absorbing medium can be written as
\begin{equation}
\frac{\partial \bm{Y_{B}}}{\partial s} \: = \: - \bm{\tilde{K}_B} \: \bm{Y_{B}} \: .
\label{absmedia}
\end{equation}
Since the statistical absorption is a symmetric operation and does not depend on the
polarization state of the mean conditional Stokes vector we can introduce a modified
conditional matrix attenuation or evolution operator
which acts on the mean conditional Stokes vector $\bm{\tilde{O}_B}(s,s')$ at the position $s'$ to 
give the mean conditional Stokes vector at the position $s$
\begin{equation}
\bm{Y_{B}}(s) \: = \: \bm{\tilde{O}_B}(s,s') \: \bm{ Y_{B}}(s') \: ,
\label{evolution1}
\end{equation}
and obeys the following limiting condition
\begin{equation}
\bm{\tilde{O}_B}(s,s) = 1 \: .
\label{limitevo}
\end{equation}
Taking the derivative of (\ref{absmedia}) and taking into account the relation
(\ref{evolution1}) a differential equation for the conditional evolution 
operator can be written as
\begin{equation}
\frac{\partial \bm{\tilde{O}_B}(s,s')}{\partial s} \: = \: \bm{\tilde{K}_B}(s) \: \bm{\tilde{O}_B}(s,s') \: .
\label{evolutiondiff}
\end{equation}
Following \citet{Lan85}, we can readily obtain
the formal solution for the inhomogeneous case (\ref{meancond3}) by a direct 
substitution, such that we can write
\begin{equation}
\bm{Y_B}(s) \: = \: \int_{s_0}^{s} \bm{\tilde{O}_B}(s,s') \bm{\tilde{j}_B}(s')  ds' + 
\bm{\tilde{O}_B}(s,s_0) \bm{Y_B}(s_0) \: .
\label{formal}
\end{equation}
where $\bm{\tilde{j}_B}(s')$ is the modified conditional emission vector (\ref{modemission}).
In the special case of a constant modified absorption matrix $\tilde{K}_B$ the 
conditional evolution operator can be written as
\begin{equation}
\bm{\tilde{O}_B}(s,s') \: = \: exp \left [ -(s-s') \tilde{K}_B \right ] \: ,
\end{equation}
where the exponential of the modified absorption matrix is given by its Taylor expansion.
Thus we see that by introducing a modified conditional evolution operator the 
formal solution for the stochastic polarized radiative transfer Eq. (\ref{meancond3}) 
is formally identical to the deterministic case \citep{Lan85}.

\subsection{Micro- and macroturbulent limits}
\label{MicroMacro}
In this section we show that the stochastic transfer equation (\ref{meancond3}) contains the 
microturbulent as well as the macroturbulent approach as limiting cases. 

We begin by considering the macroturbulent case which
follows the idea that the turbulence or the stochastic character of the 
atmosphere is exclusively perpendicular to the line-of-sight. 
Any fluctuation along the line-of-sight is by far larger than the
mean free path of the photons and is therefore negligible. On the other hand, 
the atmosphere is highly structured in the plane perpendicular to the line-of-sight. 
For observations with finite resolution, there always exists an ensemble of 
individual line-of-sights which traversing through different atmospheric regimes. 
The statistical variation within that finite resolution element is usually
described by a simple probability density function.
Macroturbulence, whether for velocity field, magnetic fields or a more general 
atmospheric vector like $\bm{B}$, certainly provides only a crude and 
rather simplistic representation of a turbulent or fluctuating atmosphere 
with large coherent structures. 
In terms of our mesostructured approach, macroturbulence can be 
described by a fluctuation rate that goes to zero or 
a correlation length that goes to infinity respectively.
If we therefore consider Eq. (\ref{meancond3}) and take the limit as 
$\gamma \rightarrow 0$ or $\lambda \rightarrow \infty$ respectively, 
we obtain the transport equation for the mean conditional Stokes vector 
$\bm{Y_{B}}$ in the macrostructured or macroturbulent limit
\begin{eqnarray}
\lim_{\lambda \to \infty} \frac{\partial \bm{Y_{B}}}{\partial s} \: = \: - \bm{K_B Y_{B}} + \bm{j_B} \: .
\label{macrolimit1}
\end{eqnarray} 
This leads to a deterministic transport equation where both statistical scattering 
terms vanish and the conditional Stokes vector in the regime $\bm{B}$ 
completely decouples from all other atmospheric components.
A final integration over the state space at the top of the atmosphere 
(Eq.\ref{meanstokes}) gives the mean observable Stokes vector for each frequency
\begin{equation}
<\bm{I}(s)> \: = \: \int \bm{Y_B}(s) \: q(\bm{B}) \: d\bm{B} \: .
\label{macrolimit2}
\end{equation}
In the case of a random macroscopic velocity field where $\bm{B} = \bf{v}$ and a 
Gaussian LOS velocity distribution p(v), this equation turns into the familiar 
(macroturbulent) convolution integral.

To derive the microturbulent limit, we have to take into account that the  
transport equation (\ref{meancond3}) was derived in the particular limit of a small ratio 
of the path element $\Delta s$ to the correlation length $\lambda$ (see Appendix \ref{Appendix_Master}) 
and therefore the limit in equation (\ref{meancond3}) cannot 
be performed directly. Instead, let us first consider the underlying Kubo-Anderson 
process Eq.(\ref{kubo}) with a depth independent fluctuation rate. If we 
take the limit as $\gamma \rightarrow \infty$ or 
$\lambda \rightarrow 0$ we arrive at the relation
\begin{eqnarray}
\lim_{\lambda \to 0} p(\bm{B}'',s+\Delta s \mid \bm{B}',s) \: =  \: q(\bm{B}'') \: .
\label{kubolimit1}
\end{eqnarray}
The transition probability from $\bm{B}'$ to $\bm{B}''$ is then exclusively 
determined by the probability of final state $q(\bm{B}'')$ and, hence, entirely uncorrelated 
from the initial state $\bm{B}'$. 
Note, that for a depth independent fluctuation rate the following 
relation holds $p_{stat} = q$, which
states that the stationary solution of the master equation (\ref{meancond3}) is given by the 
prescribed probability density $q$ \citep{Briss74}. 
The Kubo-Anderson process for this limit is therefore simply given by the overall stationary
(s-independent) probability density function $p_{stat}(\bm{B})$ of the atmosphere. 
Since the Kubo-Anderson process is now independent of the spatial variable $s$ 
it follows that the master equation (\ref{master_equation}) is
stationary as well and must be identical to zero.
As the conditional Stokes vectors are in general $\bm{Y_B} \neq 0$ this further 
implies that the sum of the second and third term on the 
r.h.s. of equation (\ref{meancond3}) must be in an equilibrium state such that both 
will mutually cancel.
This equilibrium can be expressed by the following relation,
\begin{eqnarray} 
\bar{\gamma} \int \bm{Y_{B'}} \: p_{stat}(\bm{B}') \: d\bm{B}' = \gamma \int \bm{Y_{B}} \: p_{stat}(\bm{B}') \: d\bm{B}' \: .
\label{microrelation1}
\end{eqnarray}
From Eq. (\ref{kubolimit1}) and  Eq. (\ref{modgamma}), we see that $\bar{\gamma} = \gamma$
and Eq. (\ref{microrelation1}) can be written as
\begin{eqnarray} 
\gamma \int (\bm{Y_{B'}} - \bm{Y_{B}}) \: p_{stat}(\bm{B}') \: d\bm{B}' \: = \: 0 .
\label{microrelation2}
\end{eqnarray}
From this relation it immediately follows that
\begin{equation}
 \bm{Y_{B}} \: = \: <\bm{I}> \: .
 \label{micromean}
\end{equation}
This relation is a direct consequence of the strong statistical coupling 
between the conditional Stokes vectors.
If we insert the relation (\ref{micromean}) into the transport equation 
(\ref{meancond3}) and taking into account relation (\ref{microrelation1}) 
we can write the transport equation for the mean conditional Stokes vector 
in the microturbulent limit as
\begin{eqnarray}
\lim_{\lambda \to \infty} \frac{\partial \bm{Y_{B}}}{\partial s} \: = \: - \bm{K_B}<\bm{I}> + \bm{j_B} \: .
\label{microlimit1}
\end{eqnarray}
Note, that a strong statistical coupling of the conditional Stokes vectors,
expressed by a small correlation length, leads to a rapid fluctuation of the
atmospheric conditions along the photon trajectory. 
Due to this strong coupling and high transition probabilities, 
photons will fluctuate on very small scales many times from one 
atmospheric regime to another before they 
eventually get thermalized or escape from the atmosphere. 
The high probability for a statistical absorption and scattering event
compared to the thermalization probability provides an almost 
complete mixing on smallest scales such that the
conditional Stokes vector will rapidly converge to the mean Stokes vector (see relation (\ref{micromean})).
The microturbulent transport equation for the mean unconditional Stokes vector can then be 
derived from Eq. (\ref{meanstokes}) to give
\begin{equation}
\frac{\partial <\bm{I}>}{\partial s} \: = \: 
- \int \bm{K_B}<\bm{I}> p_{stat}(\bm{B}) \: d\bm{B} + \int \bm{j_B} p_{stat}(\bm{B}) \: d\bm{B}  .
\label{microlimit2}
\end{equation}
using  $<I> \: \cong \: I$, we can write the transport equation for the mean 
Stokes vector in the microturbulent limit finally as
\begin{equation}
\frac{\partial \bm{I}}{\partial s} \: = \: <\bm{K}>\bm{I} \: + \: <\bm{j}> \: ,
\label{microlimit3}
\end{equation}

Note that, mathematically, the microturbulent limit is reached in the 
strict limit of $\lambda \rightarrow 0$ or $\gamma \rightarrow \infty$.
However, as the \emph{complete mixing} of the photons or the ratio
of statistical scattering to true absorption is in fact controlled
by the opacities in the individual atmospheric regimes 
the effective microturbulent limit will be reached 
for a finite correlation length (see Sect. \ref{Experiments}).

\subsection{Stokes profile asymmetries}
\label{Asymmetry}
The asymmetry properties of Stokes profiles originating in a stratified atmosphere 
were the subject of several studies over the recent decades 
\citep{Auer78,Lan83,SA89,Sol93a,Gro00,Steiner00,Lopez02}.
In the following, we show how the stochastic transfer equation for polarized light
can be used to describe the generation of asymmetric Stokes profiles.
We assume that the atmosphere comprises an ensemble of different magnetic regimes where
each regime may have its own thermal properties and different LOS velocities. 
For simplicity, we assume that each regime or component is height-independent 
such that each conditional absorption matrix $\bm{K}_B$ is constant 
with $s$. Note again, that the vector $\bm{B}$ contains all relevant
atmospheric parameters. According to \citet{Lan81}
for the Zeeman effect in an isolated triplet the seven entries of the absorption matrix 
possess clear symmetry properties with respect to the central line frequency or 
wavelength at any given depth. The point of symmetry of the absorption 
profiles in the absorption matrix are defined with respect to the central frequency $\nu_0$ 
of the underlying Voigt- or Faraday-Voigt profiles.
If we denote $\nu$ as the distance from the the central frequency $\nu_0$, then the 
entries of the absorption matrix $\bm{K}_{\nu_0}$ satisfy the following relations
\begin{eqnarray}
\eta_I(\nu) \: = \: \eta_I(-\nu) \nonumber \\
\eta_Q(\nu) \: = \: \eta_Q(-\nu) \nonumber \\
\eta_U(\nu) \: = \: \eta_U(-\nu) \nonumber \\
\eta_V(\nu) \: = \: -\eta_V(-\nu) \nonumber \\
\rho_Q(\nu) \: = \: -\rho_Q(-\nu) \nonumber \\
\rho_U(\nu) \: = \: -\rho_U(-\nu) \nonumber \\
\rho_V(\nu) \: = \: \rho_V(-\nu) \: .
\label{symprop}
\end{eqnarray}
This is a direct consequence of the symmetry properties of the Voigt and 
Faraday-Voigt function. The absorption profiles 
$\eta_I,\eta_Q,\eta_U,\rho_V$ are even functions regarding to the central 
frequency $\nu_0$ and the $\eta_V,\rho_Q,\rho_U,$ are odd functions.
We recall that the evolution operator in the case of a constant absorption matrix 
$\bm{K}$ \citep{Lan85} can be written as 
\begin{eqnarray}
\bm{O}(s,s') \: = \: exp^{[-\bm{K}(s-s')]} \: = \: \sum_{n=0}^{\infty} \frac{-[\bm{K}(s-s')]^n}{n!} \:.
\label{matexp}
\end{eqnarray}
Note that the evolution operator itself shares the same symmetry properties 
about the central frequency $\nu_0$ as the entries of the absorption matrix $\bm{K}$.
We can now state, that any four component vector $\bm{X}$ whose components shares the 
following symmetry properties in the frequency domain
\begin{eqnarray}
X^1_{\nu_0} \propto \eta_{I\nu_0} \nonumber \\
X^2_{\nu_0} \propto \eta_{Q\nu_0} \nonumber \\
X^3_{\nu_0} \propto \eta_{U\nu_0} \nonumber \\
X^4_{\nu_0} \propto \eta_{V\nu_0} \: ,
\label{vecprop}
\end{eqnarray}
and which is subject to a transformation by means of the evolution operator (\ref{matexp}) 
will preserve the symmetry properties. This can be verified
by a direct multiplication, and is a direct consequence
of the symmetry behavior (the even-odd relation) of the individual 
absorption profiles in the absorption matrix.
Therefore, any evolution operator (\ref{matexp}) that acts on a vector $\bm{X}$ with
the properties (\ref{vecprop}), and which shares the same central symmetry $\nu_0$ 
leaves the symmetry properties of the vector components unaltered. 
On the other hand, non-symmetric entries or entries with different points of symmetry 
in the evolution operator matrix will in general not preserve the symmetry of the 
vector $\bm{X}$.

For the sake of clarity we change the notation from a continuous state space to a discrete,
such that the individual atmospheric components $\bm{B}_k$ can be better distinguished. 
If we now take the formal solution for the mean conditional Stokes vector $\bm{Y_{B_k}}$ 
in a semi-infinite atmosphere we get
\begin{equation}
\bm{Y_{B_k}}(s) \: = \: \int_{- \infty}^{s} \bm{\tilde{O}_{B_k}}(s,s') \: \bm{\tilde{j}_{B_k}}(s')  \: ds' \: .
\label{formal-tmp}
\end{equation}
We then use the explicit form of the modified emission $\bm{\tilde{j}_{B_k}}(s')$, 
which is the sum of the thermal emission $\bm{j_B^{Therm}}$ and the inflow from 
the statistical scattering $\bm{j_B^{Scat}}$ we can write Eq. (\ref{formal-tmp}) in the form
\begin{eqnarray}
\bm{\tilde{j}_{B_k}}(s) \: = \: \bm{j_{B_k}^{Therm}}(s) + \bm{j_{B_k}^{Scat}}(s) \nonumber \hspace{2.6cm} \\
= \: \bm{j_{B_k}^{Therm}}(s) \: + \: \sum_l \bar{\gamma}(\bm{B}_k,s) \: \bm{Y_{B_l}} \: p(\bm{B_l},s) \: . 
\label{scat_term}
\end{eqnarray}
From the symmetry properties of the evolution operator $\bm{\tilde{O}_{B_k}}(s,s')$ and the
thermal emission vector  $\bm{j_{B_k}^{Therm}}$ (note, both acting in the same regime
$\bm{B_k}$) it is clear that a transformation of $\bm{j_{B_k}^{Therm}}$ by the evolution operator 
$\bm{\tilde{O}_{B_k}}(s,s')$ leaves the 
symmetry properties of $\bm{j_{B_k}^{Therm}}$ unaltered and therefore does not lead to 
any asymmetry of the conditional Stokes vector.
This simply reflects the fact that we have assumed that the absorption matrix for
each atmospheric component is constant. 
We now concentrate on the statistical scattering term and replace $\bm{Y_{B_l}}$ on the 
r.h.s. of Eq. (\ref{scat_term}) by its formal solution and then insert the expression for
the statistical scattering into Eq. (\ref{formal-tmp}) which gives
\begin{eqnarray}
\bm{Y_{B_k}}(s) \: = \: \int_{- \infty}^{s} \int_{- \infty}^{s'} \bm{\tilde{O}_{B_k}}(s,s')   
\: \sum_l \bm{\tilde{O}_{B_l}}(s',s'') \nonumber \hspace{1.7cm} \\ 
\times \: \bm{\tilde{j}_{B_l}}(s'') \bar{\gamma}(\bm{B}_k,s') p(\bm{B}_l,s'') \: ds'' ds' \: .
\label{formal-tmp2}
\end{eqnarray}
Equation (\ref{formal-tmp2}) leads to a sum of successive applications of the local evolution operators 
$\bm{\tilde{O}_{B_k}}$ and $\bm{\tilde{O}_{B_l}}$. Despite the fact
that each atmospheric component $\bm{B}_l$ has a constant absorption matrix, 
the individual atmospheric components differ in their atmospheric parameters such 
that the matrix exponentials (evolution operators) in general do not commute. 
We therefore use the Baker-Campbell-Hausdorff formula \citep{Magnus54} to express
the product of two matrix exponential which involves all higher-order terms of the commutator, 
$[\bm{A},\bm{B}] = \bm{A}\bm{B} - \bm{B}\bm{A}$,
\begin{eqnarray}
exp(\bm{A})exp(\bm{B}) \: = \: exp(\bm{A} + \bm{B} + \frac{1}{2}[\bm{A},\bm{B}] \nonumber \hspace{2.3cm}\\
+ \frac{1}{12}[A,[\bm{A},\bm{B}]] - \frac{1}{12}[B,[\bm{A},\bm{B}]] + ...) \: .
\label{BCH}
\end{eqnarray}
Using this relation we can write Eq. (\ref{formal-tmp2}) as
\vspace{0.1cm}
\begin{eqnarray}
\bm{Y_{B_k}}(s) \: = \: \int_{- \infty}^{s} \int_{- \infty}^{s'}    
\: \sum_l exp(\bm{K}_{B_k}(s'-s) + \bm{K}_{B_l}(s''-s') \nonumber \hspace{0.0cm}\\
+ \frac{1}{2}[\bm{K}_{B_k},\bm{K}_{B_l}](s'-s)(s''-s') + ...) \nonumber \hspace{0.0cm}\\
\times \: \bm{\tilde{j}_{B_l}}(s'') \: \bar{\gamma}(\bm{B}_k,s') \: p(\bm{B}_l,s'') \: ds'' ds' \: . \hspace{-0.3cm}
\label{formal-tmp3}
\end{eqnarray}
Although each individual absorption matrix $\bm{K}_{B_l}$ has its own central frequency $\nu_l$ 
(point of symmetry) due to the individual atmospheric conditions, 
the superposition and multiplication of the individual matrices will in general 
not retain or lead to a new symmetry in the compound matrix and therefore the 
resulting (combined) evolution operator will not possess any symmetry properties.
This means that the modified emission vector $\bm{\tilde{j}_{B_l}}(s'')$ in Eq.(\ref{formal-tmp3}), 
which again receives contributions from thermal emission and statistical scattering, is 
subject to a non-symmetric compound evolution operator of the form
\begin{eqnarray}
\bm{\tilde{O}_{B_{kl}}}(s,s') \: = \: \sum_l exp( \bm{K}_{B_k}(s'-s) + \bm{K}_{B_l}(s''-s') \nonumber \hspace{1.2cm}\\
+ \frac{1}{2}[\bm{K}_{B_k},\bm{K}_{B_l}](s'-s)(s''-s') + ...) \: .
\label{combevo}
\end{eqnarray}
Regardless, of the symmetry properties of the statistical scattering vector
$\bm{j_{B_l}^{Scat}}(s'')$ at $s''$ -- which itself is the result of the 
underlying mixed atmosphere -- the symmetry of the thermal emission vector 
$\bm{j_{B_l}^{Therm}}(s'')$, which shares the local symmetry properties of the regime $\bm{B_l}$,
is no longer preserved by a transformation with the compound evolution operator 
(\ref{combevo}) and thus no general symmetry property for the 
conditional Stokes vector $\bm{Y_{B_k}}$ can evolve in this type of 
stochastic atmosphere. 
So we can conclude that it is the inherent asymmetry of the compound evolution operator 
(\ref{combevo}) that prevents any propagation of symmetric profiles.

Yet another interesting fact that directly follows from Eq. (\ref{formal-tmp3}) is 
that the degree of asymmetry is  
proportional to the fluctuation rate $\gamma(\bm{B}_k)$ or inversely 
proportional to the correlation length $\lambda(\bm{B}_k)$.
This fact will be used in the following section to introduce the 
correlation length as a diagnostic parameter.

\section{Numerical experiments}
\label{Experiments}
In this Section we perform a number of numerical experiments to demonstrate 
how mesostructures, structures of finite length, 
can affect the polarized line formation and the shape of 
Stokes profiles.  
Particular emphasis in this section is given to the generation of a asymmetric Stokes 
profiles and their dependence on the mean length scales of the underlying magnetic and 
non-magnetic structures.
We show, how a wide range of structural length scales are not appropriately described 
by a microturbulent or macroturbulent approach and that significant deviations
occur compared to our more realistic and accurate mesostructured approach. 

\subsection{Numerical solution}
\label{Numerics}
Before we begin with our model calculation we give a brief outline of the numerical procedure 
we have used to solve the stochastic transport equations. The numerical solution is based on
the discrete form of transport equation for the mean conditional Stokes vector
(\ref{meancond6}) which is given as
\begin{eqnarray}
\frac{\partial \bm{Y_{B_m}}}{\partial s} \: = \: - \bm{K_{B_m} Y_{B_m}} + \bm{j_{B_m}} + 
\sum_n \bar{\gamma}(\bm{B_{n}}) \: \bm{Y_{B_n}} \: p(\bm{B_n},s)
\hspace{0.8cm} \nonumber \\
 - \sum_n \gamma(\bm{B_{m}}) \: \bm{Y_{B_m}} \: q(\bm{B_n})  \: .
\end{eqnarray}
Each of the transport equations for the mean conditional Stokes vector couples 
the 4 mean conditional Stokes parameter defined in Eq.(\ref{meancondpar}) via the 
absorption matrix.
Additionally, the transport equations are coupled via the statistical absorption 
and scattering terms to the $n$ different atmospheric regimes. 
We therefore express the whole system by means of
a $4n \times 4n$ system matrix $\bm{M}$ which gives us the following system of 
differential equations
\begin{equation}
\frac{d \bm{\hat{Y}}}{ds} \: = \: \bm{M} \: \bm{\hat{Y}} \: + \: \bm{\hat{j}} \: ,
\end{equation}
where $\bm{\hat{Y}}$ and $\bm{\hat{j}}$ are the $4n$-element Stokes vector and thermal 
emission vector. The system matrix $\bm{M}$ can be compactly written by introducing 
the local submatrices $\bm{R_i}$ and $\bm{S_i}$ to yield the following
system of differential equations
\begin{eqnarray}
\frac{d}{ds} \left ( \begin{array}{c} Y_{B_1}^{I}(s) \\
Y_{B_1}^{Q}(s) \\  Y_{B_1}^{U}(s) \\  
Y_{B_1}^{V}(s) \\   Y_{B_2}^{I}(s) \\  
Y_{B_2}^{Q}(s) \\  Y_{B_2}^{U}(s)  \\
Y_{B_2}^{V}(s) \\ . \\ . \\ . \\  Y_{B_n}^{I}(s) \\
Y_{B_n}^{Q}(s) \\   Y_{B_n}^{U}(s) \\ 
Y_{B_n}^{V}(s) \end{array} \right ) \; = \;
\left [ \begin{array}{ccccc}
\bm{R}_1 & \bm{S}_2 & \bm{S}_3 & \cdots & \bm{S}_n \\ 
\bm{S}_1 & \bm{R}_2 & \bm{S}_3 & \cdots & \bm{S}_n \\
\bm{S}_1 & \bm{S}_2 & \bm{R}_3 & \cdots & \bm{S}_n \\
 . & . & . &  & . \\
 . & . & . &  & . \\
 . & . & . &  & . \\ 
\bm{S}_1 & \bm{S}_2 & \bm{S}_3 & \cdots & \bm{R}_n \\ 
\end{array} \right ]
\left ( \begin{array}{c} Y_{B_1}^{I} \\  Y_{B_1}^{Q} \\
Y_{B_1}^{U} \\  Y_{B_1}^{V} \\  Y_{B_2}^{I} \\  Y_{B_2}^{Q} \\  Y_{B_2}^{U} \\  
Y_{B_2}^{V} \\ . \\ . \\ . \\  Y_{B_n}^{I} \\  Y_{B_n}^{Q} \\  Y_{B_n}^{U} \\  
Y_{B_n}^{V} \end{array} \right )
\; + \;
\left ( \begin{array}{c} j_{B_1}^{I} \\  j_{B_1}^{Q} \\
j_{B_1}^{U} \\  j_{B_1}^{V} \\  j_{B_2}^{I} \\  j_{B_2}^{Q} \\  j_{B_2}^{U} \\  
j_{B_2}^{V} \\ . \\ . \\ . \\  j_{B_n}^{I} \\  j_{B_n}^{Q} \\  j_{B_n}^{U} \\  
j_{B_n}^{V} 
\end{array} 
\right ) \; . \label{system}
\\ \nonumber
\end{eqnarray}
The local absorption matrix $\bm{R}_i$ of the regime $i$ is given by the following
$4 \times 4$ matrix
\begin{eqnarray}
\bm{R}_i \; = \; 
\left [
\begin{array} {cccc}
\scriptstyle{- (\eta^I_{B_i} + \frac{1}{\lambda(B_i)})} & \scriptstyle{- \eta^Q_{B_i}} & 
\scriptstyle{- \eta^U_{B_i}} & \scriptstyle{- \eta^V_{B_i}}   \\
\scriptstyle{- \eta^Q_{B_i}} & \scriptstyle{- (\eta^I_{B_i} + \frac{1}{\lambda(B_i)})} &
\scriptstyle{- \rho^V_{B_i}} & \scriptstyle{\rho^U_{B_i}}  \\
\scriptstyle{- \eta^U_{B_i}} & \scriptstyle{\rho^V_{B_i}} & 
\scriptstyle{- (\eta^I_{B_i} + \frac{1}{\lambda(B_i)})} & \scriptstyle{- \rho^Q_{B_i}}  \\
\scriptstyle{- \eta^V_{B_i}} & \scriptstyle{- \rho^U_{B_i}} &
\scriptstyle{\eta^Q_{B_i}} & \scriptstyle{- (\eta^I_{B_i} + \frac{1}{\lambda(B_i)})}
\end{array}
\right ] \; .
\\ \nonumber
\end{eqnarray}
And the local statistical scattering matrix $\bm{S}_i$ is given by
\begin{eqnarray}
\bm{S}_i \; = \; 
\left [
\begin{array} {cccc}
\scriptstyle{- \frac{p(B_i)}{\lambda_{B_i}}} & \scriptstyle{0} & 
\scriptstyle{0} & \scriptstyle{0}   \\
\scriptstyle{0} & \scriptstyle{- \frac{p(B_i)}{\lambda_{B_i}}} &
\scriptstyle{0} & \scriptstyle{0}  \\
\scriptstyle{0} & \scriptstyle{0} & 
\scriptstyle{- \frac{p(B_i)}{\lambda_{B_i}}} & \scriptstyle{0}  \\
\scriptstyle{0} & \scriptstyle{0} &
\scriptstyle{0} & \scriptstyle{- \frac{p(B_i)}{\lambda_{B_i}}}
\end{array}
\right ] \; .
\\ \nonumber
\end{eqnarray}
The system is integrated by a fourth-order Runge-Kutta method \citep{Press92} 
with appropriate boundary values for the mean conditional vectors 
$\bm{Y_{B_m}}$. The initial conditions are given according to
Eq. (\ref{icond}).
\begin{equation}
\bm{Y_{B_m}} = S_c(\tau_{i_N}) \bm{e}_0 + \left [ \frac{dS_c(\tau_{i_N})}{ds} \right ]
\bm{K}^{-1}(\tau_{i_N}) \bm{e}_0 \: ,
\label{boundary}
\end{equation}
where $\bm{e}_0$ is the unity vector and $S_c$ is the Planck function in the LTE case.
The lower boundaries are chosen for a sufficient large optical depth  
such that each of the different atmospheric regimes satisfy the condition 
$\tau_{i_N} \geq 2$.

To obtain the mean Stokes vector at the top of the atmosphere we finally
sum up the individual contributions of the mean conditional Stokes vectors 
according to Eq. (\ref{meanstokes}).

\subsection{A mesostructured magnetic atmosphere -- 2 fluctuating components}
\label{TwoCompField}

To demonstrate the importance of the correlation length
for the line formation process we begin by considering a simple model of a 
stochastic atmosphere consisting of two different types of components.
  
The arrangement of the different structures is such that we assume to have 
an ensemble of magnetic structures which are embedded in field-free regions. 
This picture is very similar to the often invoked 
two-component flux tube scenario often used for modeling unresolved magnetic 
structures in the solar network and internetwork. 
However, this scenario differs twofold,
first, the resolution element comprises a whole ensemble of many individual structures 
and is not limited to a single static flux tube with a prescribed geometry, 
and second, the length scale of the individual magnetic and non-magnetic structures 
are finite. 
In the following we prefer therefore the term ensemble rather than components to 
emphasize the intrinsic complexity of the atmosphere. 
The line-of-sight length scale of the individual magnetic and non-magnetic structures 
in our model will be determined by the same correlation length L. 

The individual magnetic structures in our model share
the same physical parameters, the same holds for the non-magnetic structures.
Note also, that we make no particular assumptions about the geometry of the 
magnetic flux structures and the ambient non-magnetic velocity field.
For simplicity reasons we assume the same temperature 
stratification for the magnetic and the field-free structures. 
This should be a reasonable choice for small scale magnetic structures 
($<$ 100 km in horizontal diameter) where thermal conduction by radiation 
efficiently smears out temperature differences. 
But even for larger structures it has no significant influence on the 
qualitative behavior of the here presented model calculations.
The underlying temperature and pressure structure is that of the quiet sun model 
atmosphere, HOLMUL, of \citet{Hol74}. 
The magnetic field strength within the magnetic structures is assumed to be
500 G (Gauss) with no internal LOS velocity. 
In the ambient medium we assume a downflow LOS velocity 
of 550 m/s to mimic a intergranular downflow. 
The total filling factor of the magnetic structures, which corresponds to the 
probability value in our stochastic scenario, is 
set to 5\%. Despite its simplicity this model should reflect 
the essential features of magnetic flux structures embedded in intergranular lanes. 
For this numerical experiment we have used the iron line 
Fe {\scriptsize I}$\lambda 6302 \AA \:$.
The coupled discrete stochastic transport equation (\ref{meancond6}) for each conditional 
Stokes vector is solved according to the numerical procedure described in Sect. 
(\ref{Numerics}). 
Note again, that we use the same correlation length L for both
ensemble structures such that L is independent of the particular atmospheric regime.
\begin{figure*}[t]
\centering
\includegraphics[width=8.5cm]{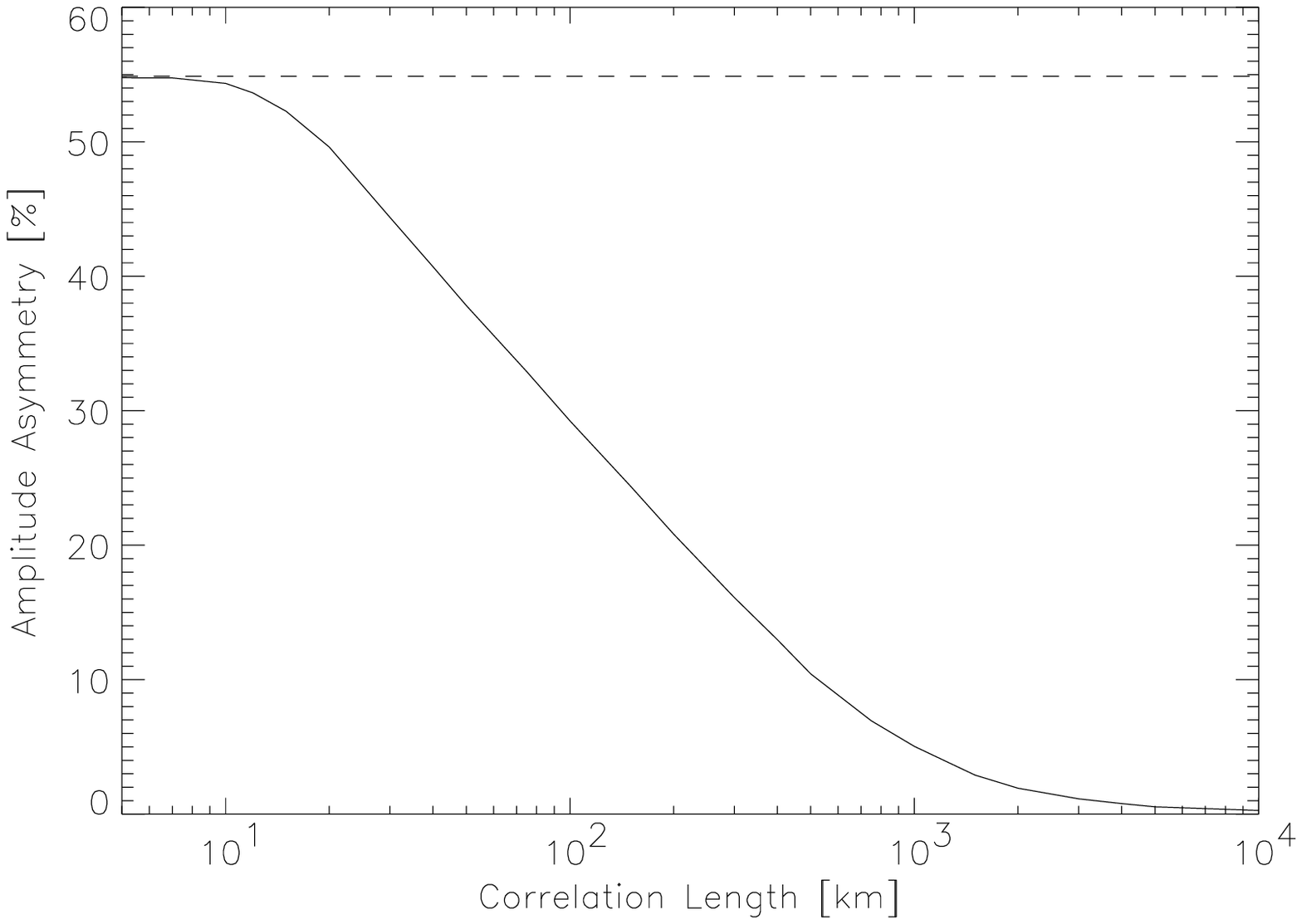}
\includegraphics[width=8.5cm]{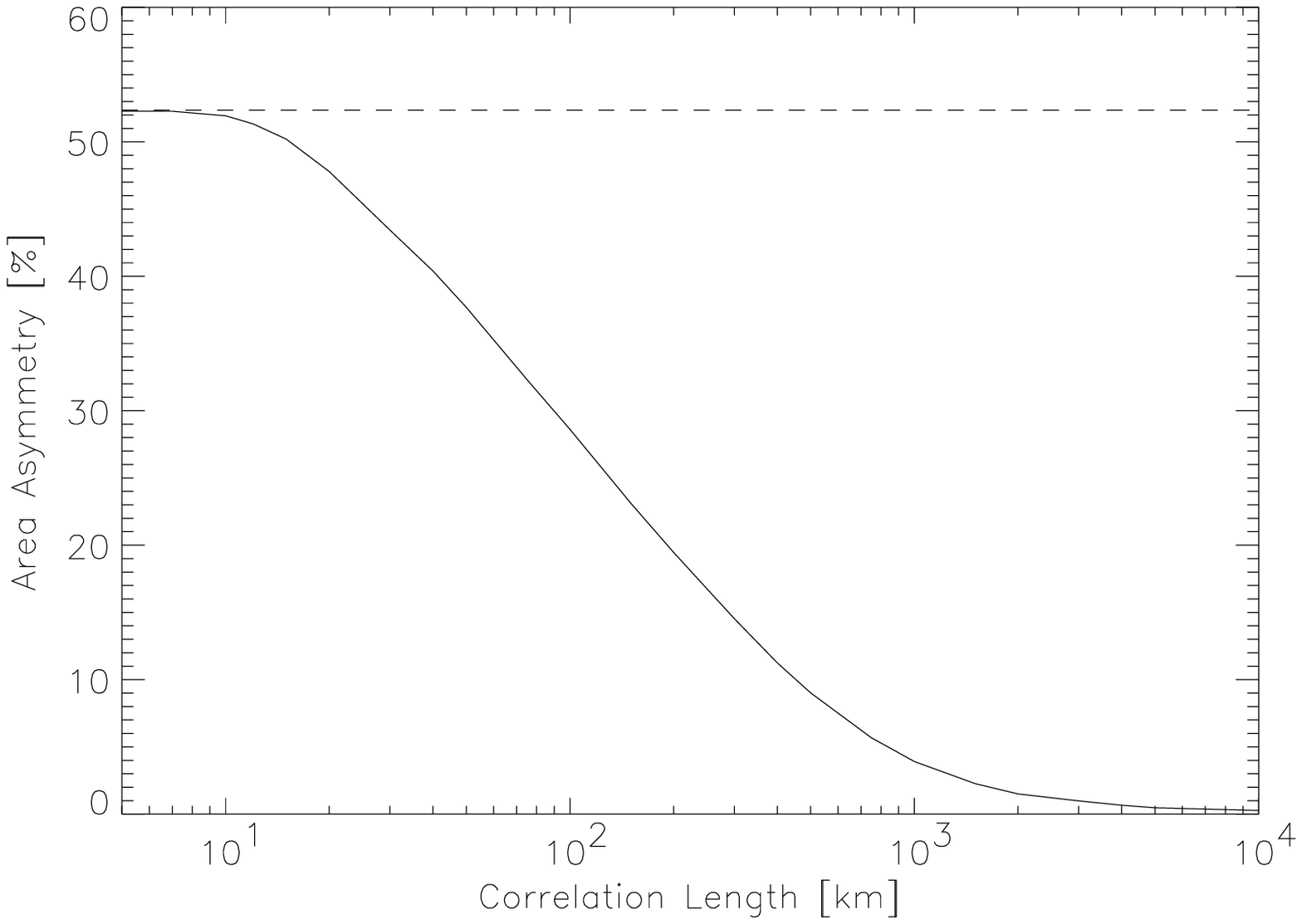}
\caption{The variation of the Stokes V asymmetry parameters (amplitude left; area right) 
as a function of the correlation length L. 
The dashed lines indicate the microturbulent limit for the asymmetry parameters calculated 
under the MISMA approximation.
Note the rapid decrease of the Stokes V asymmetry with increasing correlation length.}
\label{fig:figure1}
\end{figure*}

We now begin to change the correlation length L from very small (microturbulent) 
to very large values (macroturbulent) which results in a 
gradual increase in the (line-of-sight) order of the system (atmosphere). Such a 
decrease of the fluctuation rate should also affect the Stokes V profile asymmetry
in characteristic way, because of the distinct dependence of the asymmetry 
on the correlation length L (see Sect. \ref{Asymmetry}). 
To quantify the degree of Stokes V amplitude and area asymmetry we adopt the usual
definition for the amplitude asymmetry $\delta a$, 
\begin{equation}
\delta a \: = \: \frac{|a_b| - |a_r|}{|a_b| + |a_r|} \: ,
\end{equation}
and the area asymmetry $\delta A$,
\begin{equation}
\delta A \: =  s_b \: \frac{\int_{\Lambda} V d\lambda}{\int_{\Lambda} |V| d\lambda} \: ,
\end{equation}
where $a_b$ and $a_r$ are the extrema of the blue and red lobes of the Stokes V signal and 
$s_b$ the sign of the blue lobe. The integration for the area asymmetry $\delta A$ 
is performed over the entire wavelength range $\Lambda$ of the considered Stokes V signal.

\begin{figure}
\centering
\scalebox{0.245}{\includegraphics{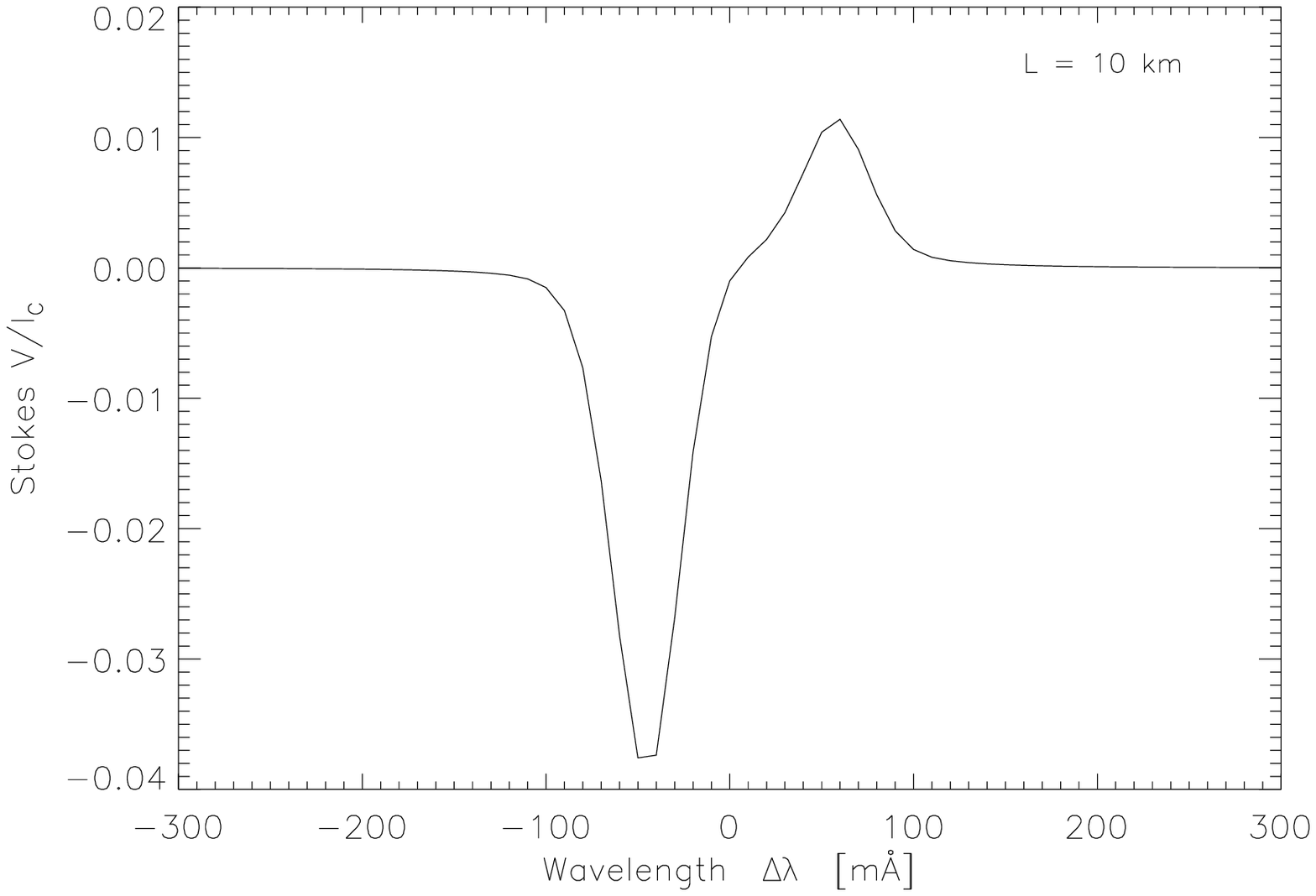}}
\scalebox{0.245}{\includegraphics{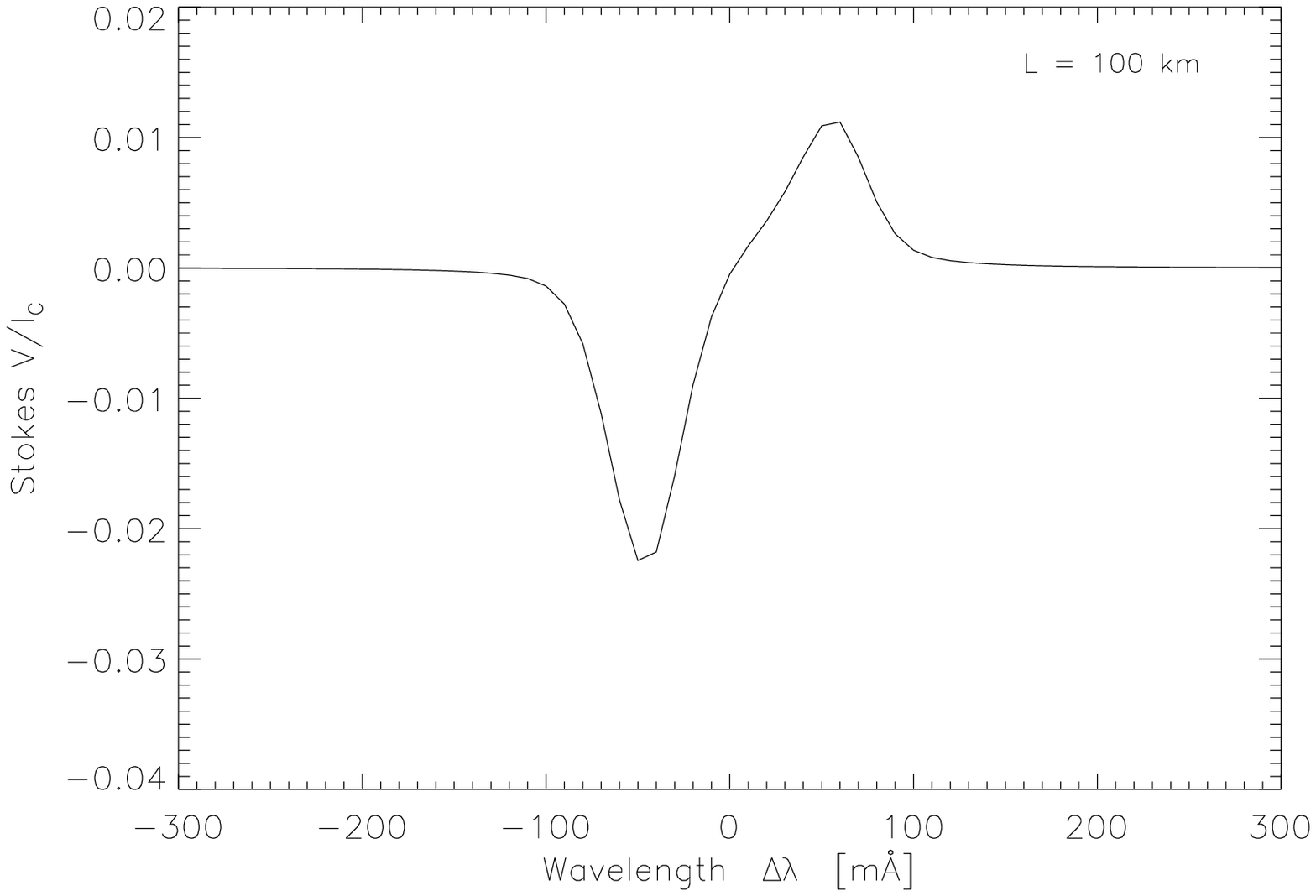}}
\scalebox{0.245}{\includegraphics{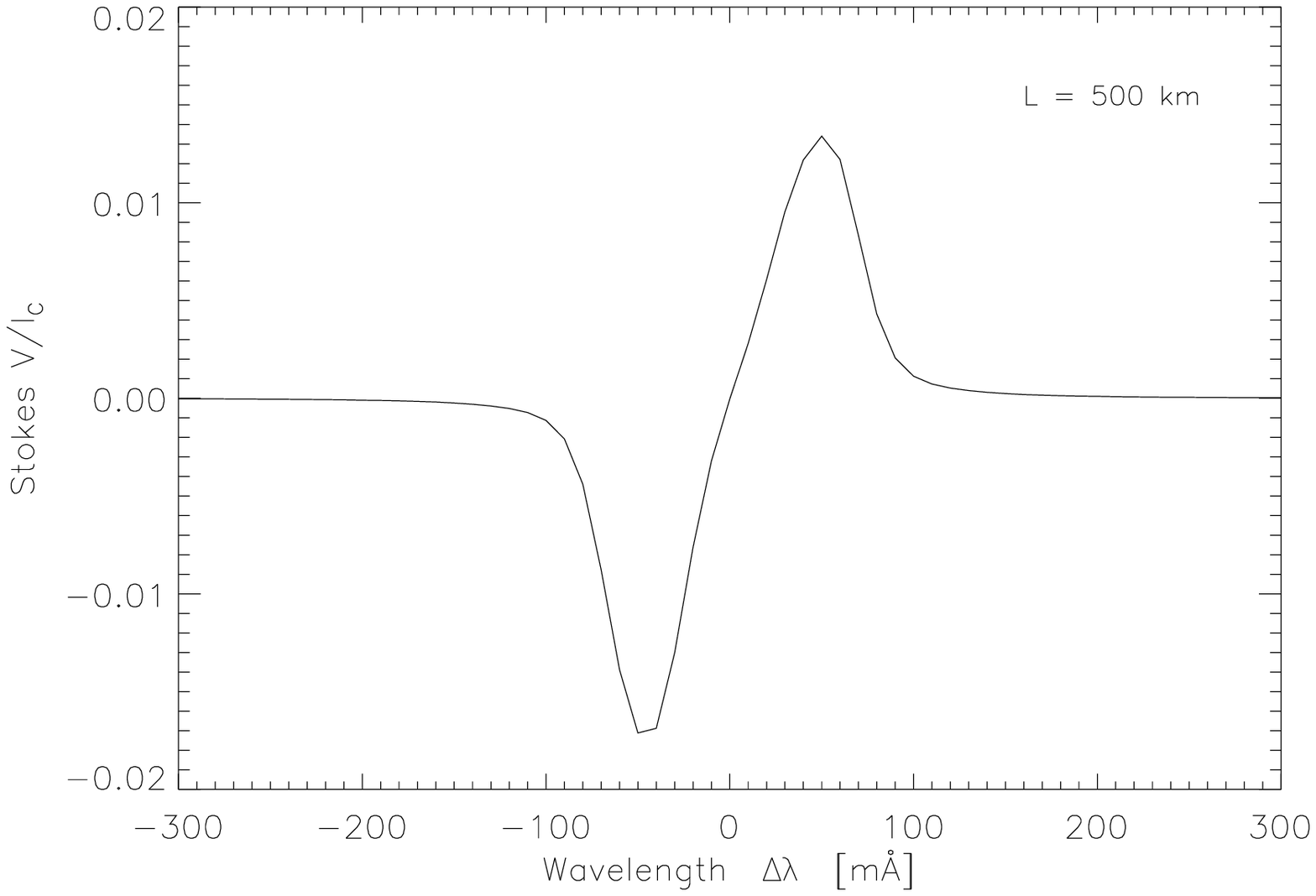}}
\scalebox{0.245}{\includegraphics{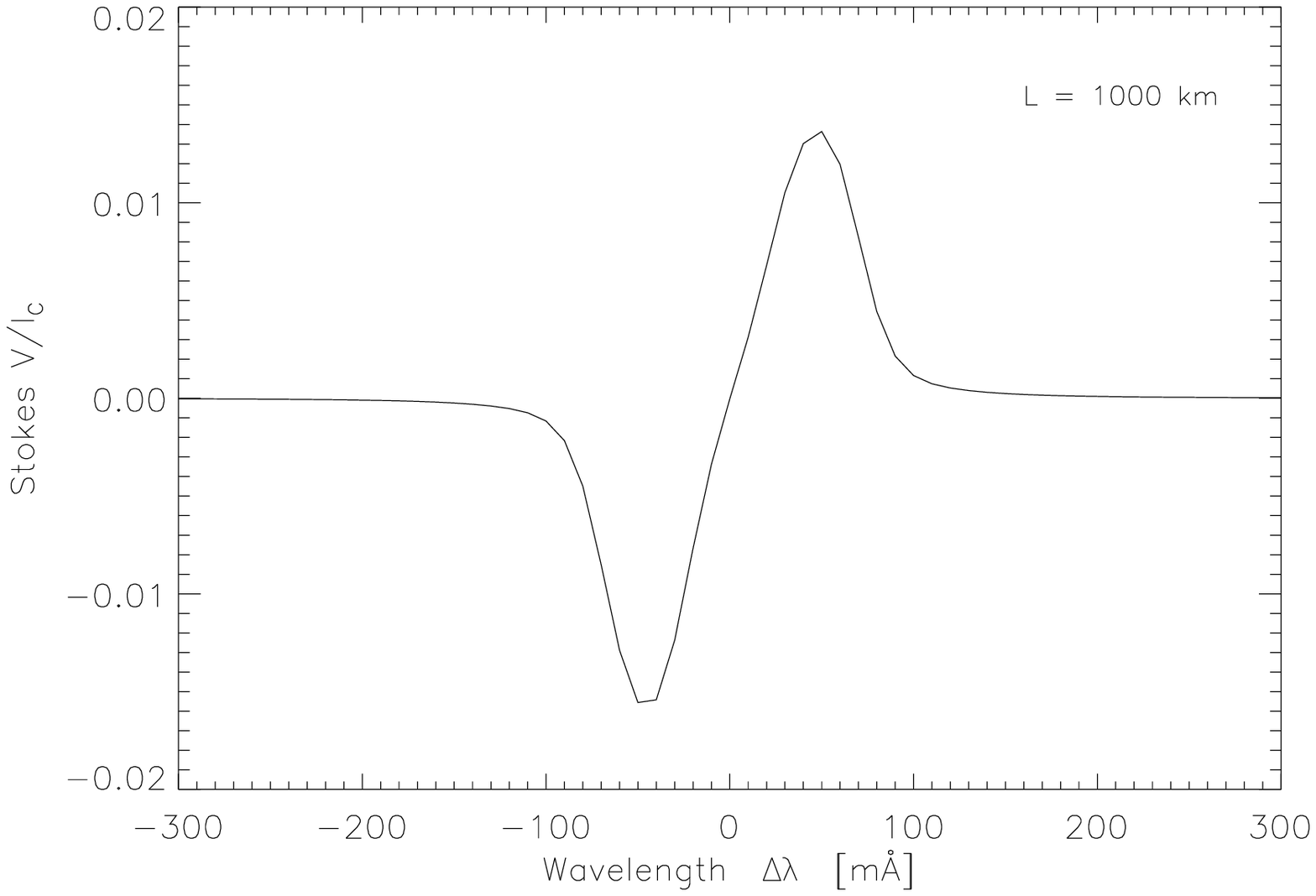}}
\caption{Sample of Stokes V profiles for a correlation length L = 10 km (upper left), 
L = 100 (upper right), L = 500 (lower left) and L = 1000 km (lower right). Note, 
the drastic changes in profile shape as well as in area and amplitude 
asymmetry.}
\label{fig:figure2}
\end{figure}

Figure \ref{fig:figure1} now shows how the amplitude and 
area asymmetry of the Stokes V profile varies with the correlation length L
of the magnetic and non-magnetic structures.
For the broad range of meso-scales (20 - 1\,000 km) we see the expected rapid decrease of 
the asymmetries with increasing correlation length L. 
This rapid decay already starts at a correlation length of approximately 15 km.
For very small correlation lengths, the asymmetries (amplitude and area) saturate 
and begin to converge into the microturbulent limit. 
For a better comparison we have also calculated the expected amplitude and area asymmetries in the 
microturbulent limit, where we have used the MISMA approximation \citep{SA96}
(marked by the dashed lines in Fig. \ref{fig:figure1}).  
For very large correlation lengths ($>$ 1\,000 km), however, the asymmetries
asymptotically converge to zero where they eventually reach -- for a correlation 
length of approximately 10\,000 km -- their macroscopic limits 
(complete symmetry). 
Fig. \ref{fig:figure1} also reveals that unlike the mathematical limits,
the microturbulent and macroturbulent limits are already reached numerically for
finite length scales. 
Figure \ref{fig:figure2} shows a sample of Stokes V profiles from our model calculations 
and displays the variation of the Stokes V profile shape with the correlation length L. 
Note, in particular, the drastic change between the Stokes V profiles calculated for 
the L = 10 km and L = 100 km case.

\subsubsection{About the microturbulent limit of photospheric structures}
\label{ReallyMicro} 
Photospheric structures below 100 km are often considered to be optically thin and 
therefore have been treated in the microturbulent limit. 
From the latter section, however, we have seen that for a varying correlation length 
the shape, as well as the Stokes V profile asymmetries, undergo a rapid change,
in particular within the first 100 km.
When there is a preferred internal alignment e.g. by buoyancy forces 
the characteristic length scales of the magnetic structures in the horizontal and 
vertical can differ by more than an order of magnitude. 
This raises the question, to what extent
structures below 100 km (measured along the LOS) can really be 
treated as microturbulent in (polarized) radiative transfer calculations ?

\begin{figure}   
\centering
\resizebox{\hsize}{!}{\includegraphics{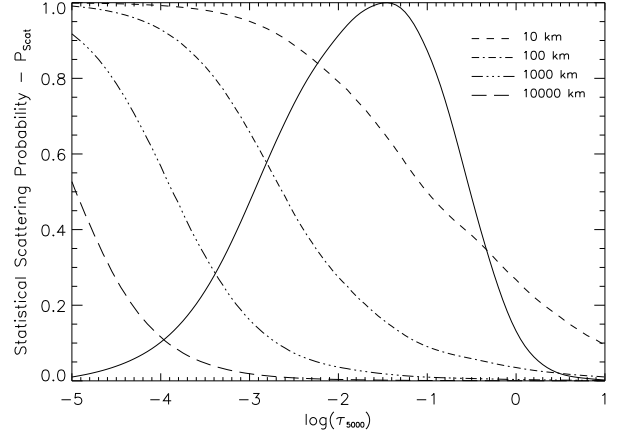}}
\caption{The statistical scattering probability
for different correlation lengths over the optical depth.
In the background, the wavelength integrated mean line depression contribution function 
for the circular polarization is given (see explanation and definition in text).}
\label{fig:figure3}
\end{figure}

From Fig. \ref{fig:figure1} we see that for a structural length scale of 
approximately 15 km the atmosphere is effectively in a microturbulent state
but apparent deviations from this microturbulent limit already occur from 20 km upward. 
For a characteristic length scale of just 100 km both 
asymmetries (area and amplitude) have lost already about 
50 \%  of their initial (microturbulent) values.
The rapid decrease of the area and amplitude asymmetries can be
qualitatively understood from the inverse dependence of the asymmetries on 
the correlation length (see Sect. \ref{Asymmetry}). 
The magnitude of the asymmetries is a direct consequence of the ratio of 
thermal absorption and statistical scattering. To quantify the
degree of fluctuation and the statistical scattering occurring in the atmosphere
in comparison to (regular) absorption processes we define the statistical scattering 
probability as follows
\begin{equation}
P_{scat} \: = \: \frac{\kappa_{stat}}{\kappa_c + \kappa_l + \kappa_{stat}} \: ,
\end{equation}
where the statistical scattering term $\kappa_{stat}$ is the reciprocal of the correlation 
length L, $\kappa_c$ the continuous absorption coefficient at line center and 
$\kappa_l$ the line absorption coefficient. 
We have calculated the statistical scattering probabilities for a number of different
correlation lengths, where we have used the same atmospheric condition as in the previous
section.
In Fig. \ref{fig:figure3} we can see the run of the statistical scattering probability
over the logarithm of the optical depth for different correlation lengths. 
For a better illustration and a better estimate of the line forming regions
we have plotted in the background of 
Fig. \ref{fig:figure3} the wavelength integrated mean line depression contribution function 
for the circular polarization (solid line). 
The wavelength integrated mean line depression contribution function 
for the circular polarization is an integral measure 
that assess the contributing layers of the emergent circular polarized signal in a 
stochastic atmosphere. The exact definition is given in Appendix \ref{Appendix_CFRCP}.

In very deep layers ($log(\tau_{5}) \ge 0$) of the atmosphere we see from 
Fig. \ref{fig:figure3} that for all correlation lengths but the smallest 
(10 km, the effective microturbulent limit) the 
probability that photons are statistically scattered into other regimes is very small.
Most of the photons in these deep layers are subject to true absorption processes and
will be thermalized before they can scatter into other atmospheric regimes. 
With increasing geometrical height or decreasing optical depth we 
see a gradual increase of the statistical scattering probability for all curves. 
But note that the increase of the scattering probability and the slope of the 
curve have a distinct dependence on the correlation length. 
In very high layers of the atmosphere the optical depth is so small that 
almost all the photons are subject only to statistical scattering regardless of their
correlation length.
Very interesting in that context are the statistical
scattering probabilities of the L = 10 km (short dashed line) case and the 
L = 100 km (dashed dotted line) case.
The two slopes significantly differ from each other even though both may be
regarded in classical terms as microturbulent.
The L = 10 km curve for the effective microturbulent case already indicates a 
significant scattering probability for very deep layers, such that the entire
line profile (wings and core) is affected by the strong statistical scattering.
Whereas, the L = 100 km curve indicates a rather small scattering probability for 
the deepest line forming layers around $log(\tau_{5}) = 0$. 
At an optical depth where the integrated Stokes V contribution function reaches its maximum,
the L = 10 km curve has a value of approximately 70 \%  which means that
photons are much more likely to encounter a transition to another atmospheric regime (70 \%)
than being thermalized in their current regime. 
Thus, most of the photons are statistically scattered many times before 
they are finally thermalized in a true absorption process. 
At the same optical depth the scattering
probability for the L = 100 km case is approximately 25 \% and therefore it has a 
less pronounced dependence on the statistical scattering which results 
in a decrease of the line asymmetry of approximately 50 \%.
 
This behavior, even for a relatively moderate downflow velocity of 550 m/s,
as assumed here, demonstrates that the microturbulent or microstructured (MISMA) 
approach does not give reliable results for atmospheric structures which 
exceeds the effective microturbulent limit.

\subsection{A magnetopause in a mesostructured atmosphere}
\label{MagPause}

In the following, we want to take a closer look on the polarized line formation 
in an atmosphere that comprises a magnetopause like boundary.
A typical example for such a scenario might be encountered in the vicinity of a rapidly
expanding large flux tube structure \citep{Steiner00}.
Similar structures with large overlaying canopies could be recently observed in 
magnetohydrodynamic simulations \citep{Schaf05} where these 
canopies can cover large parts of the adjacent granular cell.
The scenario we have assumed here for our model calculation is intended to be a crude 
approximation of a possible situation encountered in the photospheric internetwork. 
The approximation we consider here can be conceived as an ensemble of flux tube like
structures which are in pressure equilibrium with their field-free surroundings. 
These structures rapidly expand until the individual structures finally merge together 
at a certain geometrical height to form a volume filling magnetic atmosphere.
The salient feature of this model is the variation of the characteristic size of 
the individual magnetic structures with height such that these structures evolve 
from small to large scale structures.

Again, we assume a simple 2-ensemble atmosphere with a number of magnetic structures 
which all shares the same physical conditions as well as a number of non-magnetic structures.  
Furthermore, we assume that the magnetic structures are in pressure equilibrium with the 
non-magnetic surroundings, we therefore write the following relation for the height 
dependence of the magnetic field strength
\begin{equation}
B(z) \: \simeq \: B(z=0) \: exp(-z/H_B) \:,
\label{fieldstrat}
\end{equation}  
where $H_B$ is the magnetic field strength scale height which is twice the gas 
pressure scale height and $B(z=0)$ the magnetic field strength at the geometrical
height $z=0$. For $H_B$ we have assumed a value of $H_B = 250$ km. 
\begin{figure*}[t]
\centering
\includegraphics[width=8.5cm]{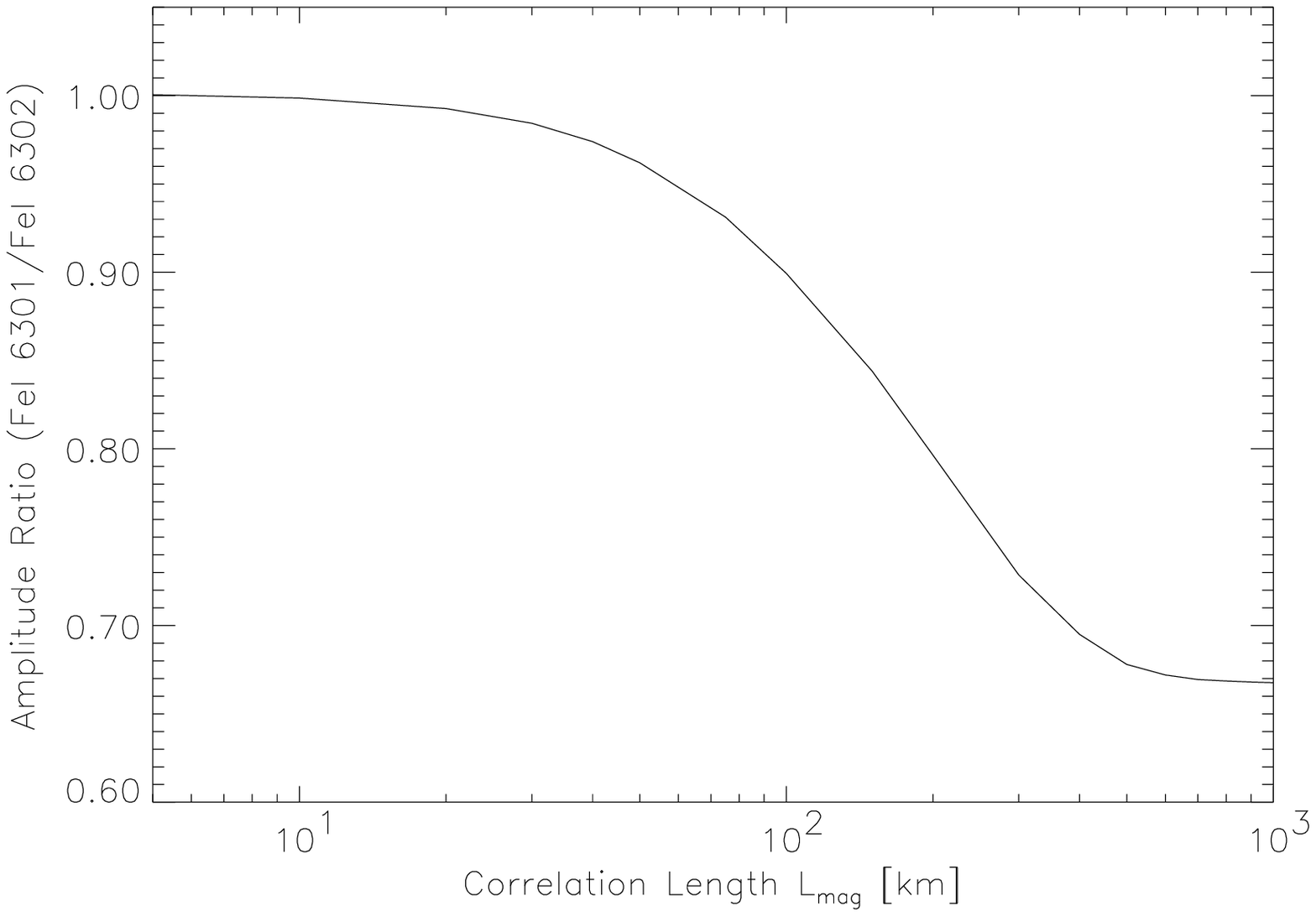}
\includegraphics[width=8.5cm]{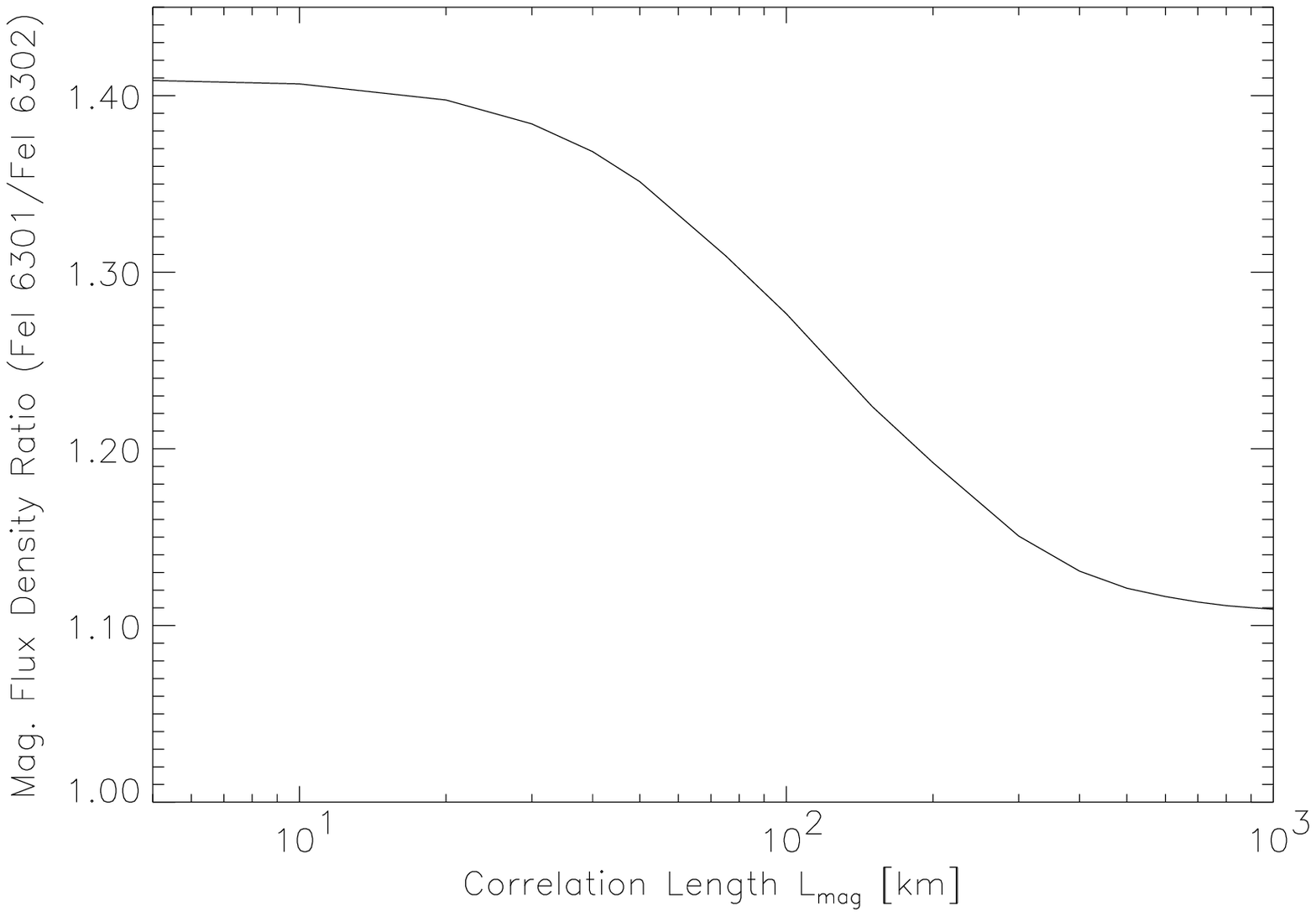}
\caption{The amplitude ratio $a^{6301/6302}$ vs. the correlation length (left)
and The ratio of the longitudinal magnetic flux densities 
$f^{6301/6302}$ vs. the correlation length (right).
The amplitude ratio as well as the magnetic flux ratio
exhibit a clear dependence on the underlying length scale of the magnetic structures.}
\label{fig:figure4}
\end{figure*}
The magnetic ensemble structures in our model scenario are mainly 
vertically orientated and we have adopted a magnetic field strength at the base $z=0$ of 
$B(z=0) = B(\tau = 1) = 500$ G. 
For clarity and to focus on the effect of a varying correlation length, we 
have ignored any possible velocities in the magnetic and non-magnetic structures.
Again, we have used the temperature stratification of the HOLMUL atmosphere 
for both ensemble structures. 
The merging height of the canopies (the magnetopause) is assumed to be at a
geometrical height of $z \approx 450$ km 
which corresponds to a logarithmic optical depth of $log(\tau_{5}) \approx -3.65$ 
in the non-magnetic HOLMUL atmosphere. 
At this height the field strength in the magnetic structures has already dropped 
to a value of approximately 80 G. 
For the magnetic ensemble structures we assume a fixed probability value of 1 \% 
at $z=0$ .
The non-magnetic ensemble structures have a probability value of 99 \% at $z=0$ 
and we assume a fixed quasi macrostructured correlation length of
L = 1000 km below the magnetopause for these structures. 
A large correlation length seems to be justified because of the much larger fill fraction 
of the non-magnetic structures. 
To mimic the rapid transition for the majority of the line-of-sights 
at the boundary of the magnetopause from whereon the atmosphere is assumed to be 
fully magnetized, we set the probability value for the magnetic structures to 
100 \% and their correlation length to L = 1000 km from $log(\tau_{5}) \approx -3.65$ on upward.

For this scenario, we have calculated synthetic Stokes spectra for the two neutral 
iron lines Fe {\scriptsize I} $\lambda 6301 \AA \:$ and Fe {\scriptsize I} $\lambda 6302 \AA \:$.
This is a prominent line pair which is often used for Zeeman diagnostics. 
Our particular interest here is to demonstrate how this prominent iron line pair 
depends on the length scale of the atmospheric structures below the magnetopause. 
For this reason we take a closer look on the differential behavior
of the Stokes V profiles of the two iron lines. A good measure to assess the strength
of the profiles is the amplitude ratio 
or the line-ratio method \citep{Stenflo94} which is commonly used to distinguish 
if the observed Stokes profiles are the result of an unresolved 
weak (sub-kG) or strong (kG) magnetic field.
The amplitude ratio is defined as $a^{6301/6302} = a(V^{6301})/a(V^{6302})$ , 
where a(V) represents the amplitude of the respective Stokes V profile.
Yet, another method going in the same direction to distinguish weak from strong magnetic field
fields is the ratio of the longitudinal
magnetic flux density derived from the two iron lines $f^{6301/6302} = B_{Long}^{6301}/B_{Long}^{6302}$
\citep{Dom03a}.
For the magnetic flux density $B_{Long}$ we have used the following approximation \citep[see][]{Socas04b}
\begin{equation}
B_{Long} \: \simeq \: \frac{C}{g_{eff}} \: \frac{\int_0^{\infty} V(\lambda) \: d\lambda}
{\int_0^{\infty} [ dI(\lambda)/d\lambda] \: d\lambda} \: ,
\label{beffective}
\end{equation}  
where C is a constant with $C= -1/(4.67 \times 10^{-13} \lambda^2)$ and $g_{eff}$ is the 
effective Land$\acute{e}$ factor of the line.
The relation (\ref{beffective}), the flux ratio $f^{6301/6302}$ as well as the 
amplitude ratio $a^{6301/6302}$ are direct consequences of the weak field 
approximation \citep{Stenflo94}.

Keeping all the atmospheric parameters fixed, except the correlation length 
of the magnetic structures below the magnetopause we have calculated the 
synthetic line profiles of the two iron lines for a series of correlation lengths.
In Fig. \ref{fig:figure4}  we have plotted the 
amplitude ratio $a^{6301/6302}$ and the longitudinal magnetic flux density ratio 
$f^{6301/6302}$ over the correlation length of the magnetic structures. 
At first glance we see a rather surprising result, 
despite the fact that the magnetic field strength is intrinsically weak $(B(z = 0) = 500$ G)
(and even rapidly decreasing with height) the amplitude as well as the flux density 
ratio suggest for small correlation lengths a strong underlying magnetic field in the 
kilo-Gauss range.
The ratio for the Stokes V amplitudes shows a value of 1.0 for small correlation lengths,
although we would expect for our intrinsically weak field that the amplitude ratio
is similar to the ratio of the corresponding Land$\acute{e}$ factors of the two 
iron lines which is approximately 0.67 \citep{Sol93b}.
Moreover, one would not expect a dependence on the underlying length scale.
As the magnetic flux density is also retrieved from the weak field approximation we 
also expect for the flux density ratio a value close to unity regardless of the 
correlation length. 
Our results show that this is not the case. For a small correlation length
the flux density ratio converges to a value of 1.41 while
for a correlation length, L = 1000 km, a flux density ratio of 1.1 is obtained.
It is obvious that with increasing correlation length the 
amplitude as well as the magnetic flux density ratio decrease. 
For a correlation length of approximately 300 km the value of the amplitude ratio 
and the flux density ratio converge to their (expected) weak field values. 
On the other hand, we see that for structures which have correlation lengths smaller 
than 100 km significant deviations from the expected values occur. 
Once again, we see that the underlying correlation length 
-- the characteristic length scale of the magnetic structures -- 
has a decisive impact on the polarized radiative transfer and the 
resulting Stokes V profiles.
\begin{figure}
\centering
\scalebox{0.245}{\includegraphics{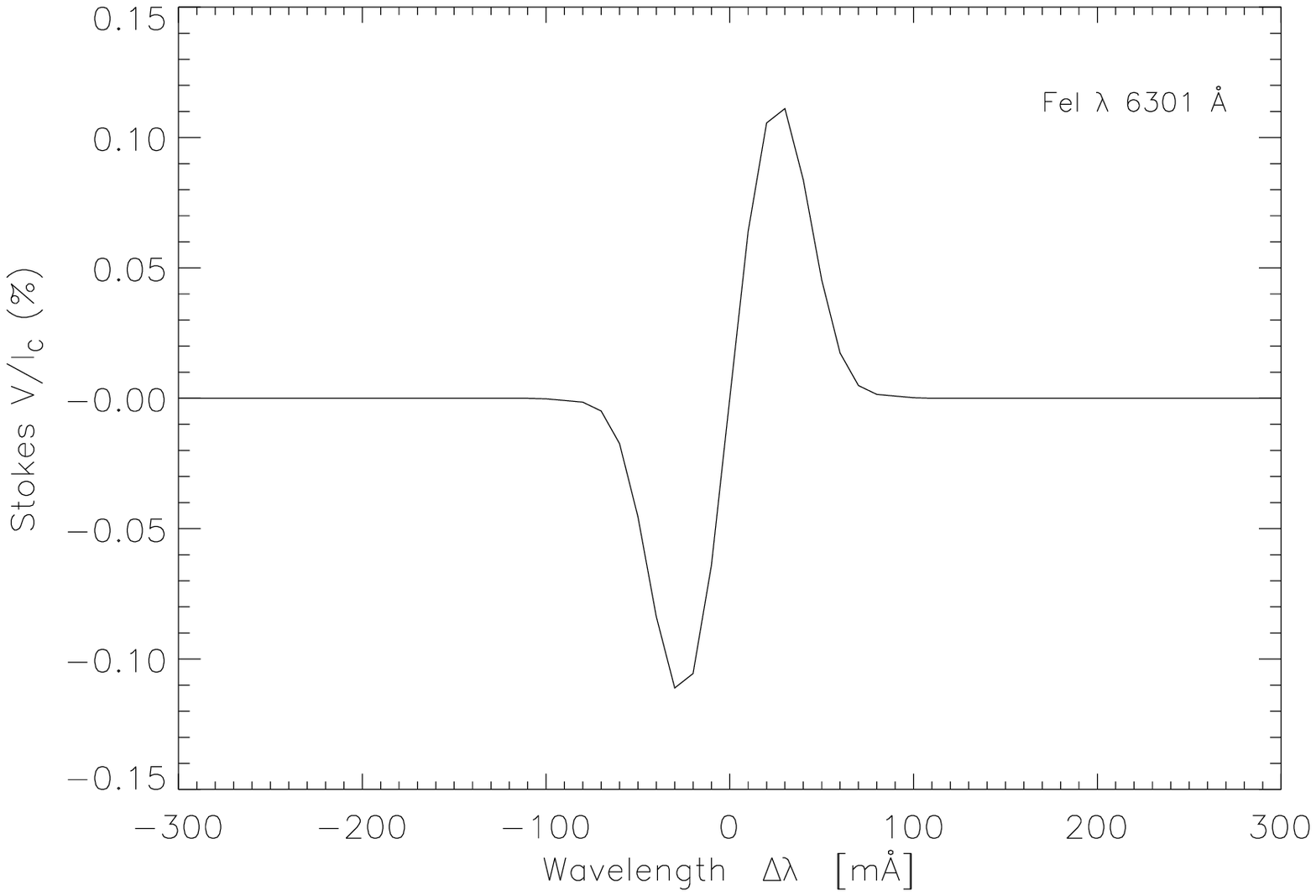}}
\scalebox{0.245}{\includegraphics{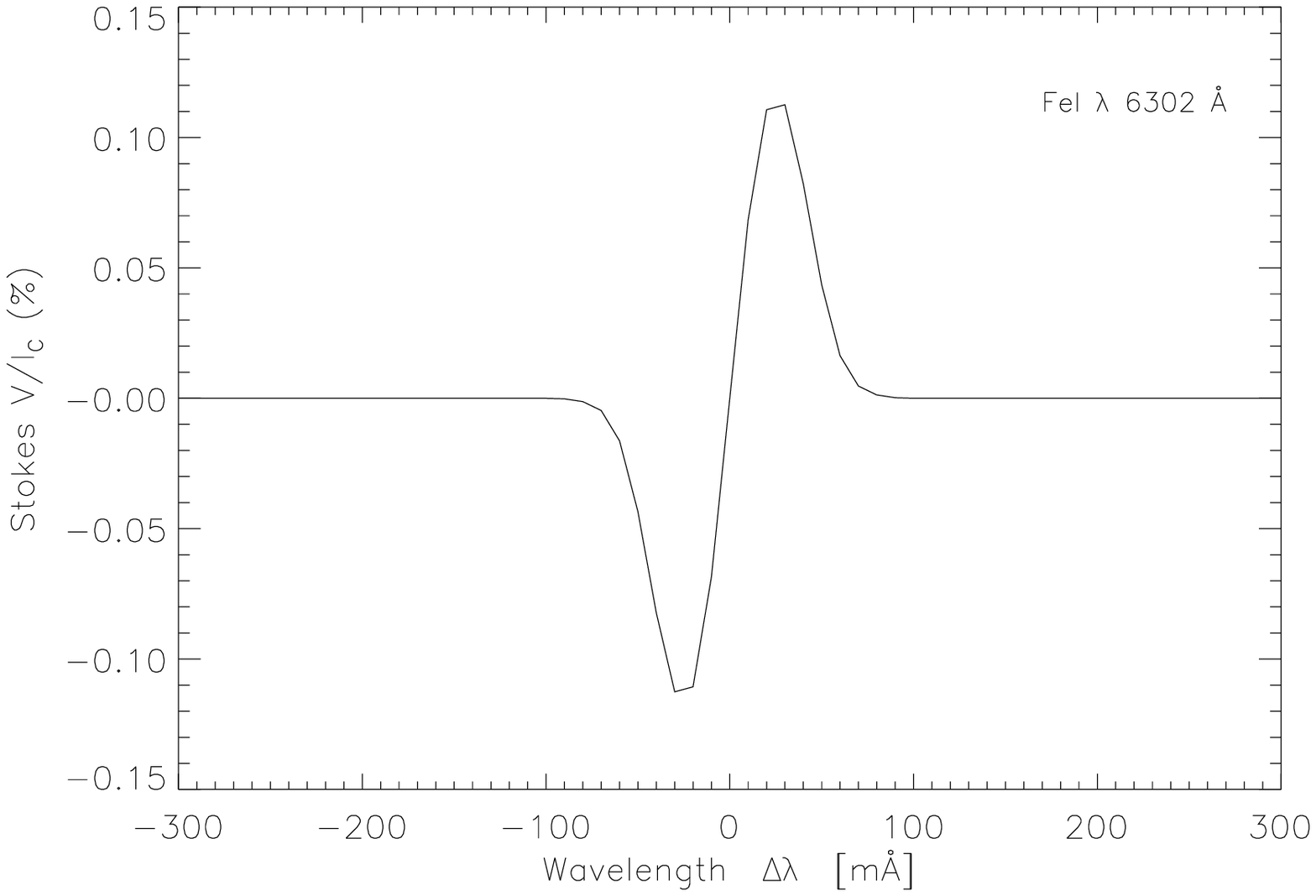}}
\scalebox{0.245}{\includegraphics{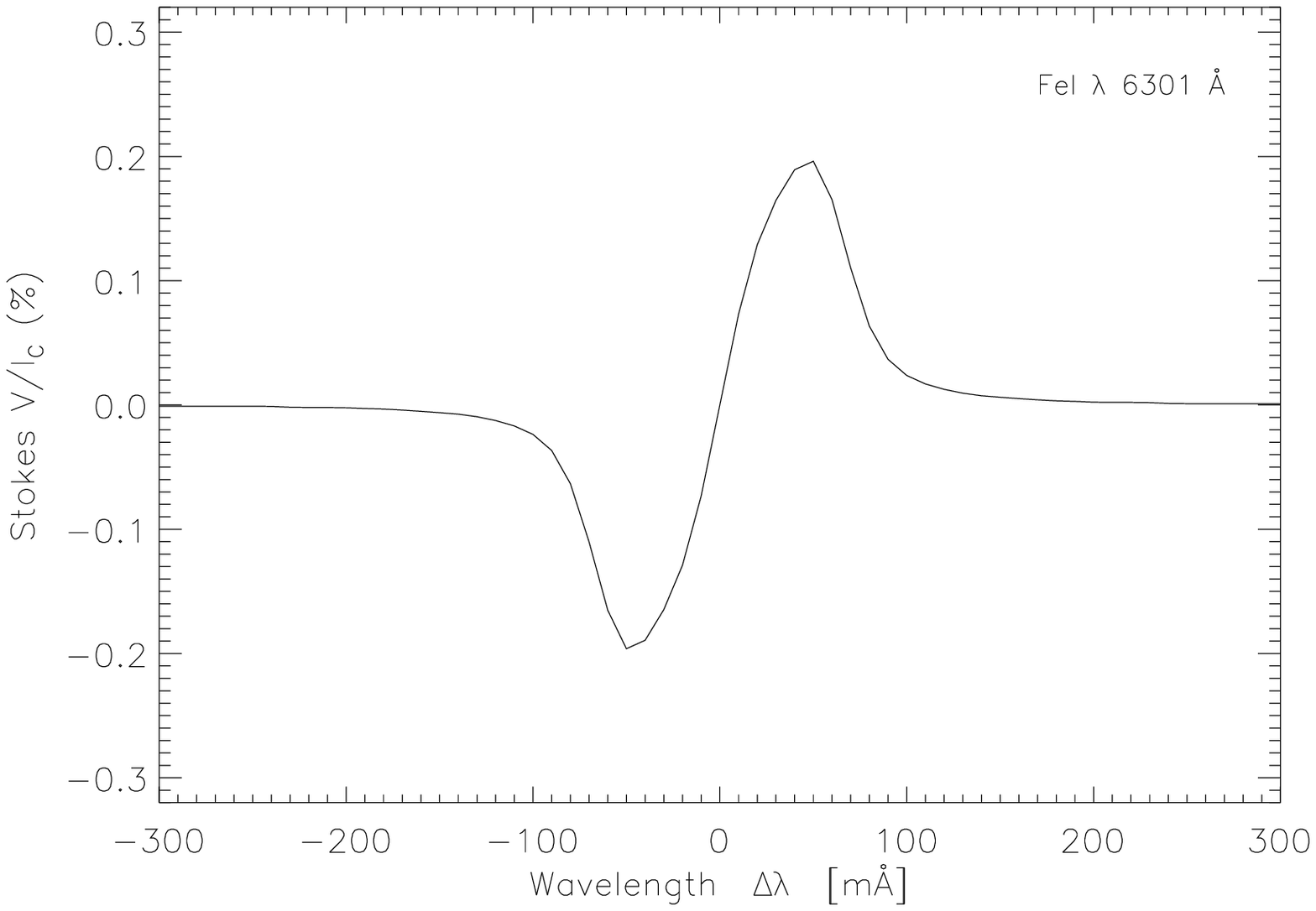}}
\scalebox{0.245}{\includegraphics{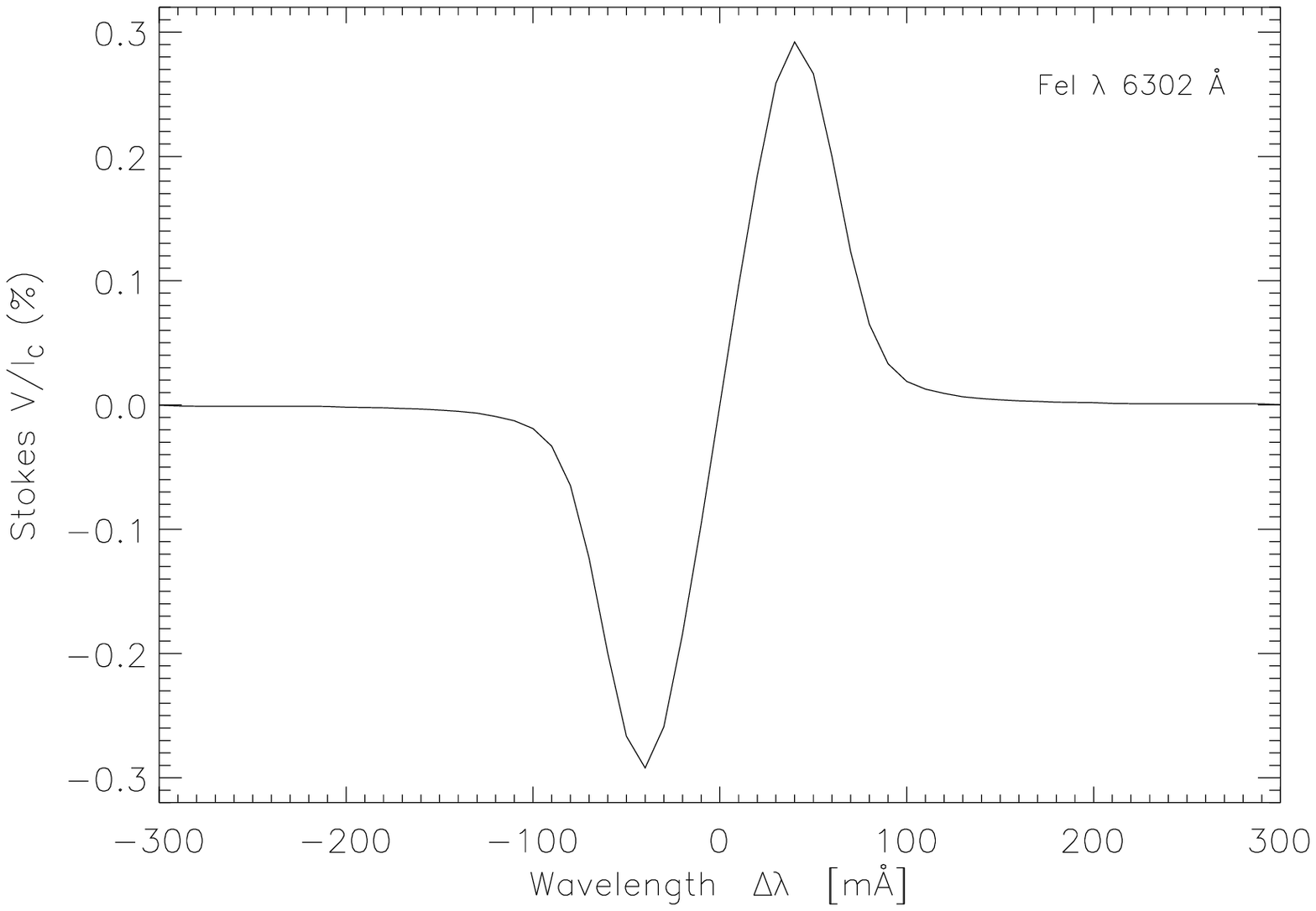}}
\caption{The upper left and right display the Stokes V profiles for the 
 Fe {\scriptsize I} $\lambda 6301 \AA \:$ (left) and  Fe {\scriptsize I} $\lambda 6302 \AA \:$ (right) 
 line calculated with a correlation length of L = 10 km. 
 Both profiles in the L = 10 km scenario have the same amplitude and therefore would
 falsely indicate the presence of strong magnetic fields. 
 The lower profiles which show the case for L = 1000 km, exhibit the
 expected behavior of the profile amplitudes for an intrinsically weak field.}
\label{fig:figure5}
\end{figure} 
To better illustrate the drastic profile variation we have plotted 
the Stokes V profiles for the two iron lines Fe {\scriptsize I} $\lambda 6301 \AA \:$ 
and Fe {\scriptsize I} $\lambda 6302 \AA \:$ in Fig. \ref{fig:figure5} for a 
correlation length L = 10 km (upper row) and a correlation length L = 1000 km (lower row).
This counterintuitive behavior of the amplitude and flux density ratios
can be understood in terms of the MESMA approximation. 
Figure \ref{fig:figure6} shows the wavelength
integrated mean line depression contribution function of the Stokes V profiles 
$\tilde{C}_{R;CP}$ (see Appendix (\ref{Appendix_CFRCP})
for the two iron lines.
These stochastic contribution functions $\tilde{C}_{R;CP}$ tell us where in the 
fluctuating atmosphere the main contributions to the Stokes V signals originate.
We immediately recognize the distinct shift in the contributing layers of the Stokes V signal. 
Figure \ref{fig:figure6} reveals that for a characteristic length scale of 
L = 10 km both iron lines receive their major contributions predominantly from 
above the magnetic canopy which is located at $log(\tau_{5}) \approx -3.65$. 
Whereas, for a mean length scale of L = 1000 km the main contributions
almost exclusively come from below the magnetopause.
The reduced contribution in the lower layers for both iron lines in the L = 10 km 
scenario, which is even more pronounced for the Fe {\scriptsize I} $\lambda 6302 \AA \:$ line, 
is a direct consequence of the absorption of polarized line photons in the 
non-magnetic layers. The non-magnetic structures do not only fill a 
substantial \emph{horizontal} fraction of 99 \% of the resolution volume, they also have, 
and this is even more important, a larger correlation length and therefore once 
photons are statistically scattered into these field-free structures, they have only 
a very small chance of being backscattered into a magnetic structure. 
Structures with large correlation length are only in weak statistical contact with 
their surrounding. Whether photons are being scattered into, or originate from 
a non-magnetic structure, the likelihood for staying in the non-magnetic regime for the
rest of their trajectory is much higher for the photons than making another transition 
to a magnetic structure. 
As the scattering probability is very high in the magnetic structures
(for low correlation lengths), many of the photons carrying the 
polarized information are scattered into the non-magnetic regimes where a
substantial number of them is absorbed due to the increased opacity in the non-magnetic structures.
This absorption of polarized photons in the non-magnetic structures is the main
cause for the reduction of the contribution in lower layers. Beginning from the 
magnetopause the atmosphere is magnetically coherent (essentially macro-structured) 
with a probability value of 100 \% and a correlation length of L = 1000 km. 
Although the contribution to the Stokes V signal from these high layers is very small in absolute terms,
it is still significant compared to the strongly reduced contribution from below the magnetopause. 
Hence, the absorption below the magnetopause is so efficient that the 
small contributions from above the magnetopause outweighs the contribution from below. 

The situation is quite different for magnetic structures with larger correlation lengths.
Although the probability value of the magnetic structures 
below the magnetopause is still only 1\% the enhancement of the correlation length 
leads to a decrease of the statistical scattering from the magnetic structures into the non-magnetic 
ones. Hence, less photons are being scattered into the denser non-magnetic structures 
and more polarized line photons are able to stay for their entire trajectory through the atmosphere 
in the same magnetic structure. 
The increasing coherency of the individual magnetic structures leads to the more 
\emph{regular} shaped form of the contribution functions (normal line formation) 
for the L = 1000 km case. 
\begin{figure*}[t]
\centering
\includegraphics[width=8.5cm]{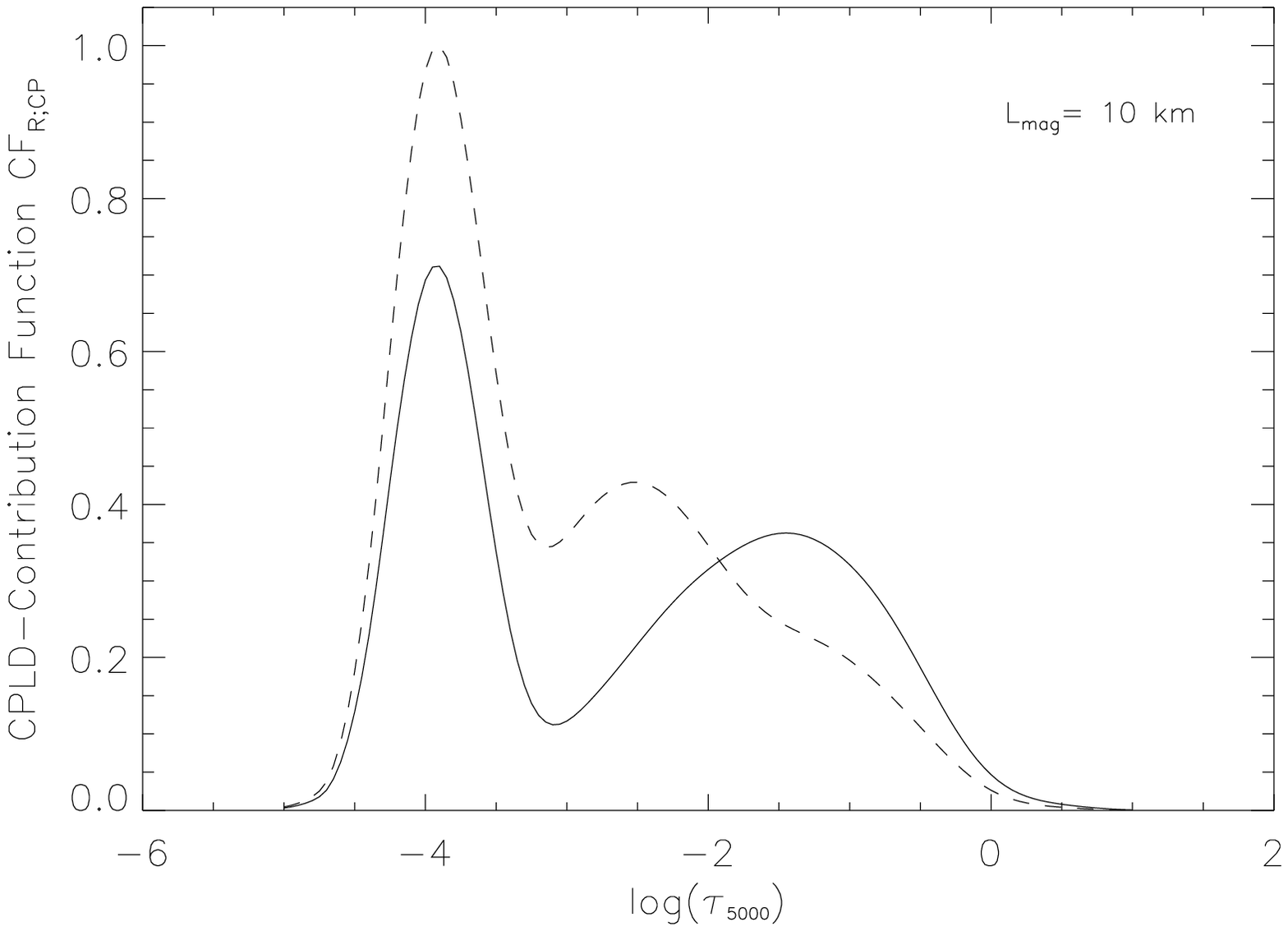}
\includegraphics[width=8.5cm]{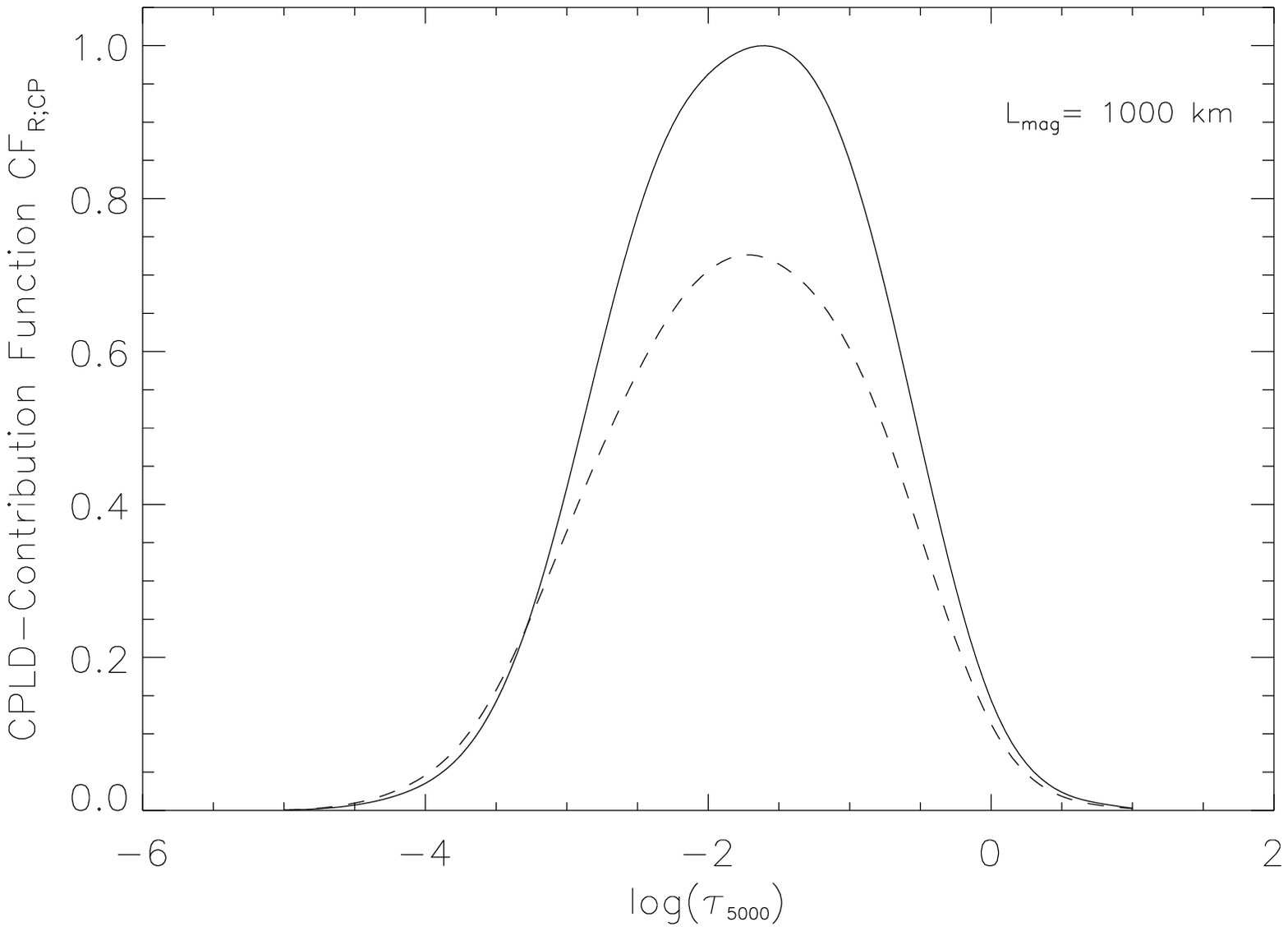}
\caption{The wavelength integrated mean line depression contribution function over the optical 
depth for the Fe {\scriptsize I} $\lambda 6301 \AA \:$ (dashed lines) and Fe {\scriptsize I} 
$\lambda 6302 \AA \:$ line (solid lines) for a correlation length of L = 10 km (left) and for
L = 1000 km (right). Note, the drastic changes of the contribution functions for both lines 
simply by varying the correlation length.
For a correlation length of L = 10 km the contributions to the Stokes V signal 
come predominantly from above the magnetopause. Note, the enhanced contribution for the 
Fe {\scriptsize I} $\lambda 6301 \AA \:$ line in the L = 10km case (left).
In the L = 1000 km case (right) one sees that for both iron lines the main contributions
come from below the magnetopause.}
\label{fig:figure6}
\end{figure*}
But, what is the cause of the unusual line ratios for small correlation lengths ? 
Now, as we have understood why 
there is such a significant difference in the contributing layers with respect to 
the correlation length, let us turn to the question of what causes the small
but decisive difference in the contribution functions for the two iron lines.
From Fig. \ref{fig:figure6}, we see that, although the two contribution functions 
for the two iron lines in the L = 10 km case show qualitatively the same 
behavior, there is a notable difference between both contribution functions.
The contribution function for the Fe {\scriptsize I} $\lambda 6301 \AA \:$ line 
(dashed line) shows a larger contribution in the layer above the magnetopause than that
of the Fe {\scriptsize I} $\lambda 6302 \AA \:$ line (solid line). 
As both lines in the L = 10 km case receive their main relative contributions from above the 
magnetopause the formation height of the two iron lines play an even more important role
than in the unstructured and coherent (e.g. macrostructured) case.
The larger contribution of the  Fe {\scriptsize I} $\lambda 6301 \AA \:$ line is 
exclusively due to its higher formation height compared to the 
Fe {\scriptsize I} $\lambda 6302 \AA \:$ line. 
This difference in the formation height which was also illustrated
and explained in terms of response functions by \citet{Mart06} as well as 
the above explained pronounced importance of the contributions from the higher 
layers above the magnetopause, lead to the effect that the Stokes V signal of the 
Fe {\scriptsize I} $\lambda 6301 \AA \:$ line receives an absolute 
larger contribution than the Fe {\scriptsize I} $\lambda 6302 \AA \:$ line. 
As the characteristic length scale of the magnetic structures grows 
this effect loses its significance because the peak of the
contribution function begins to shift downward below the magnetopause
and reduces the overemphasized difference in the formation height. 
This demonstrates once more that the analysis of magnetic field strengths
based on the iron line pair at 6300 $\AA$ in a structured magnetic atmosphere 
must be regarded with caution, as was also pointed out recently by \citet{Mart06}.

\subsection{A quiet sun magnetic field strength distribution}
\label{FieldDist}
 
In the recent years the possible role of the so called internetwork magnetic field 
has become increasingly appreciated. 
Despite its elusive character, the ubiquity of these weak flux fields could 
be observed in a number of high spatial and high sensitivity 
spectropolarimetric measurements \citep{Sig99,Lin99,Sig01,Lites02,Dom03a,Kho03,Socas04b}. 
The significance of these fields for the solar atmosphere
is not yet clear, but the mere fact that these fields cover most of the solar surface 
makes them probably an important ingredient for the structuring and dynamics of the 
higher layers of solar atmosphere \citep{Schri03,SA04}. 
The characterization of the internetwork magnetic field in terms of empirical as well as 
statistical parameters from Hanle and Zeeman effect measurements is currently under debate. 
Recently \citet{Dom06} proposed a set of empirical 
probability density functions which describe the distribution of magnetic field strength 
in the internetwork. 
They emphasize that field strength in the kilo-Gauss range, even though they have low probabilities, 
play the dominant role in transporting most of the magnetic 
energy into the upper solar atmosphere.

For empirical estimates of these pdf's from spectropolarimetric observations 
it is important to realize that line formation is an inherent three dimensional process and 
it is therefore crucial to take into account the horizontal as well as the vertical
structuring of the atmosphere.  
In the following simple numerical model calculation we want to use the lognormal probability 
distribution function proposed by \citet{Dom06}
to demonstrate that the structural length scale of the individual
magnetic structures has an important impact on the resulting Stokes signal and that
empirical probability density functions are strongly biased by the underlying
model assumptions.

\begin{figure}   
\centering
\resizebox{\hsize}{!}{\includegraphics{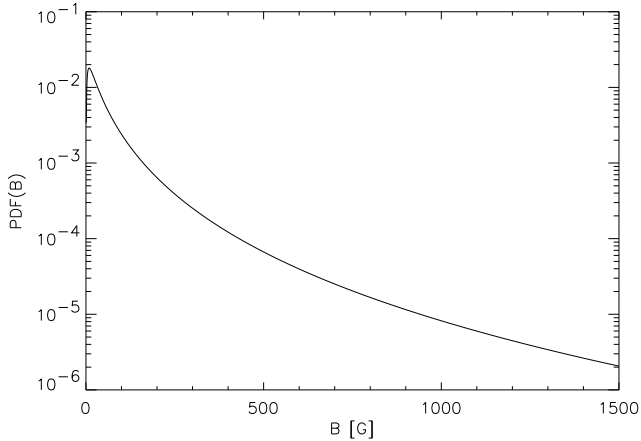}}
\caption{Empirical probability density function for the magnetic field strength.}
\label{fig:figure7}
\end{figure}

The following probability density for the magnetic field strength is used
\begin{equation}
p(B) \: = \: \frac{1}{\sqrt{\pi} \: \sigma B} \: exp \left [- \frac{(\ln B - \ln B_0)^2}{\sigma^2} \right ]
\label{empprobdist}
\end{equation}
The parameters $B_0$ and $\sigma$ are related to the first and second order moments respectively.
We have adopted the values $\sigma = 1.7$ and $B_0 = 38$  G \citep[see][]{Dom06}.
The probability density function (\ref{empprobdist}) for the magnetic field strength 
is shown in Fig. \ref{fig:figure7}, note,  that the 
probability for a field strength larger than 1000 G is only approximately 0.33 \%.
It is now crucial to realize that this empirical probability density function comprises 
a broad range of different field strength regimes, a weak and probably purely 
turbulent range as suggested by Hanle effect measurements, 
a wide range of intermediate field strengths which are approximately in equipartition 
to their non-magnetic surroundings and a strong field range which is assumed 
to be populated by small intense kilo-Gauss structures. 
This picture is also supported by an increasing evidence for the existence of a small scale 
mixture of weak, intermediate and strong fields in a single resolution element 
\citep{Socas04a}.

But do these different field regimes all have the same characteristic length scale ?
As magnetic buoyancy keeps stronger magnetic field structures more aligned to the vertical
direction than weaker field structures, the extent of the line-of-sight which 
traverses through an individual magnetic structure cannot be the same for all 
field strengths. There must be a dependence of the correlation length on the field strength.
To demonstrate the importance of this effect, we assume the following simplified scenario:
for field structures below one kilo-Gauss field strength we adopt a correlation length 
of 10 km (effectively microturbulent) and for structures with field strength larger 
than one kilo-Gauss we vary the correlation length in a range from 10 
to 10\,000 km.
Again, we have synthesized the two iron lines at $\lambda 6300 \AA \:$ 
and have used the temperature stratification of the HOLMUL atmosphere.
To assess and quantify the influence of the correlation length on the Stokes V profiles,
we have used the longitudinal magnetic flux density (\ref{beffective}), which 
gives us an estimate of the magnitude of the circular polarized signal.

\begin{figure*}[t]
\centering
\includegraphics[width=8.5cm]{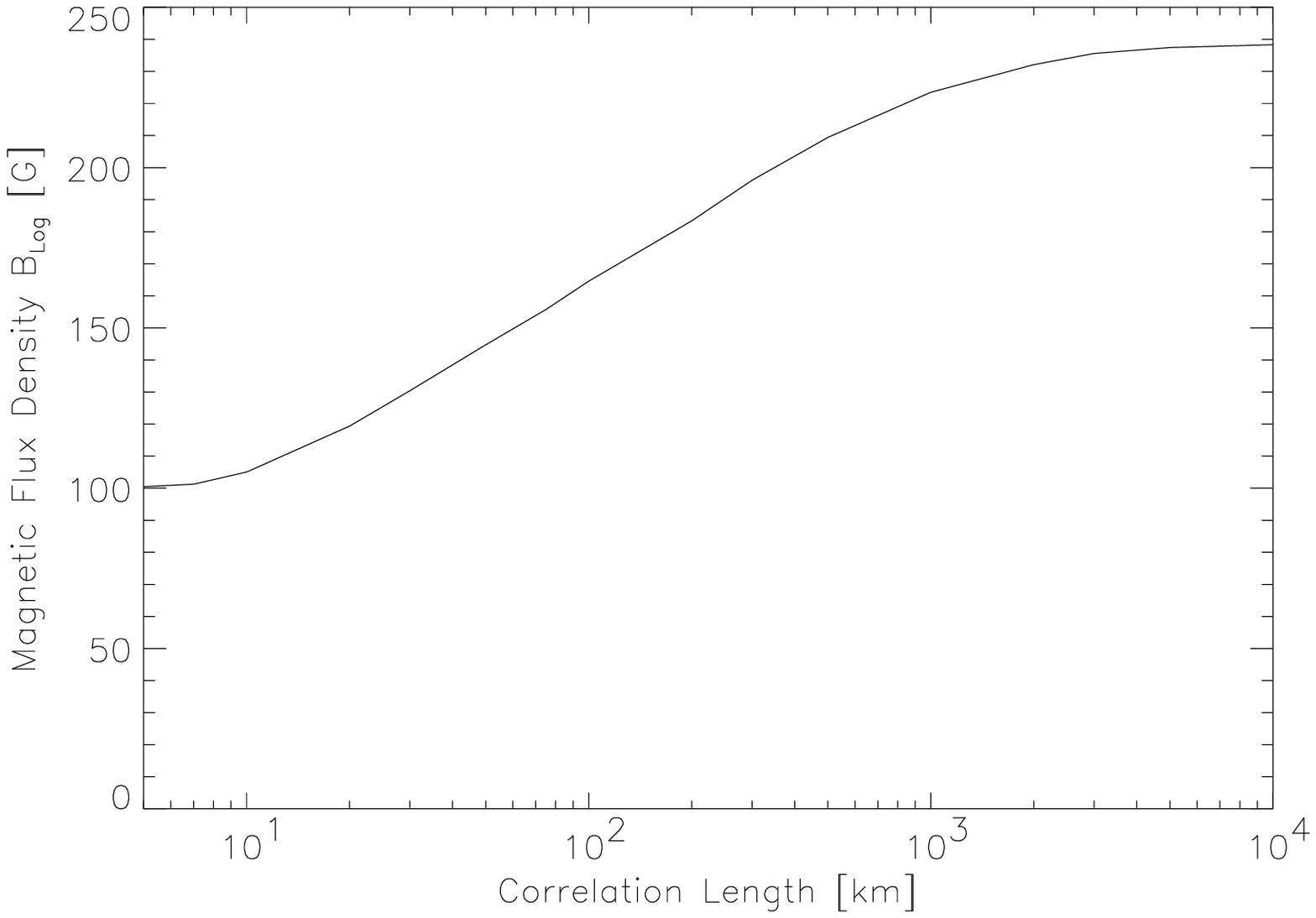}
\includegraphics[width=8.5cm]{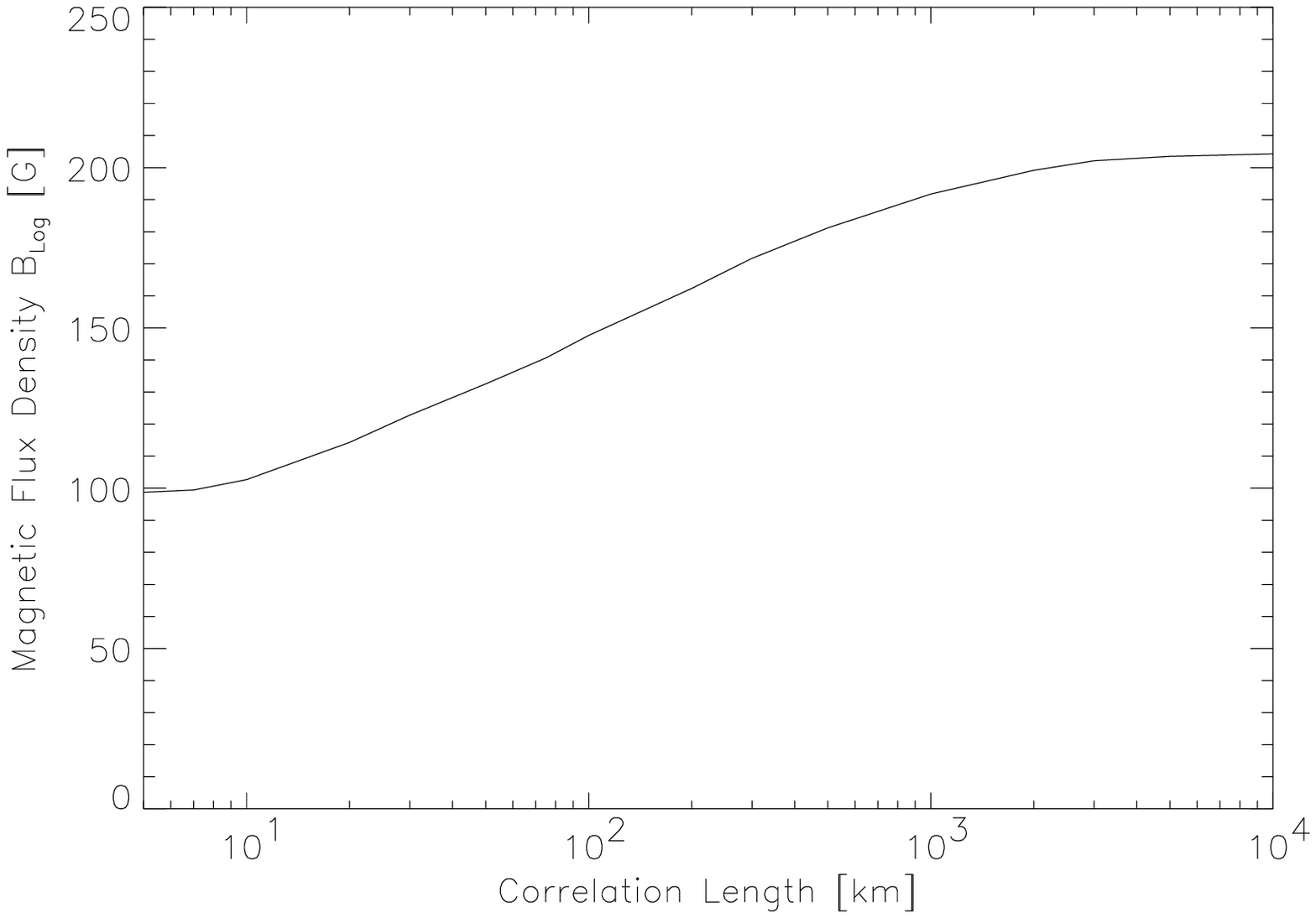}
\caption{The longitudinal magnetic flux density as a function of the strong field correlation length, 
derived from the iron line Fe {\scriptsize I} $\lambda 6301 \AA \:$ (left) and from the 
Fe {\scriptsize I} $\lambda 6302 \AA \:$ line (right).}
\label{fig:figure8}
\end{figure*}

Figure \ref{fig:figure8} shows how 
the magnetic flux densities vary with the correlation length of the
strong field structures. For both iron lines
one can see from Fig. \ref{fig:figure8} the rapid increase of the flux density with increasing 
correlation length. Please note, that no other parameter but the correlation length 
of the kilo-Gauss structures has been changed in our scenario, the probability
density function in our calculation is for all correlation lengths the same.  
For very small structures ($\sim$ 10 km) the magnetic flux density saturates and 
converges into the microturbulent limit which corresponds to the value calculated 
for the probability density function (\ref{empprobdist}) under  MISMA approximation. 
For very large structures ($\geq$ 1000 km) the value for the
magnetic flux density converges into the macroturbulent limit. 
We see that just by increasing the correlation length of the strong field structures 
from micro-scales to macro-scales the magnetic flux densities 
(and, hence, the circular polarizations of both iron lines)
are more than doubled.
This, again, demonstrates how drastic the underlying length scale 
of the magnetic structures affects the process of polarized line formation. 
Although the underlying probability density function 
(\ref{empprobdist}) has not changed, the effect of the strong field structures 
is more and more pronounced simply by increasing their correlation lengths. 
The growing correlation length is responsible for the additional magnetic flux
derived from both spectral lines.

\begin{figure}   
\centering
\resizebox{\hsize}{!}{\includegraphics{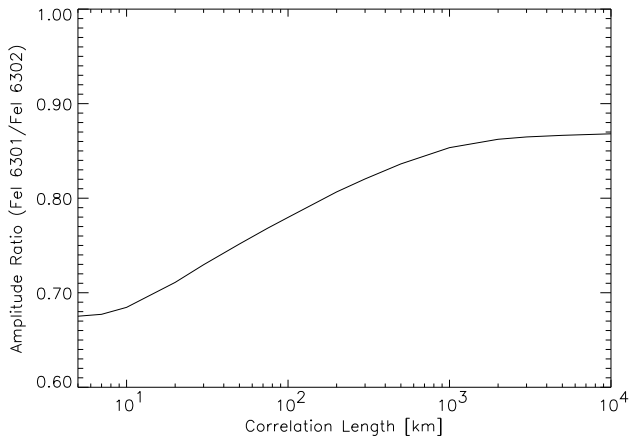}}
\caption{The amplitude ratio between the two iron lines 
Fe {\scriptsize I} $\lambda 6301 \AA \:$ and  Fe {\scriptsize I} $\lambda 6302 \AA \:$
as a function of the strong field correlation length. 
Note, that the underlying pdf is for all correlation 
lengths the same.}
\label{fig:figure9}
\end{figure}

Another intriguing aspect 
is the increasing difference of the flux densities calculated from both iron lines with
increasing correlation length (Fig. \ref{fig:figure8}).
The flux density for the iron line Fe {\scriptsize I} $\lambda 6301 \AA \:$ seems
to grow faster than that of the Fe {\scriptsize I} $\lambda 6302 \AA \:$ line.
This difference in the magnetic flux density ratio would suggest that the underlying 
probability density is dominated by the kilo-Gauss structures. 
This misleading effect is also highlighted by Fig. \ref{fig:figure9} which shows 
the amplitude ratio $r^{6301/6302}$ of the two iron lines as a function of the strong field 
correlation length. The amplitude ratio would also lead to the erroneous conclusion that the 
underlying probability density is strongly biased towards kilo-Gauss structures, even though
the kilo-Gauss structures make up less than 1\% of the overall distribution.

So we see again the decisive role of the correlation length and
that the simple derivation of empirical probability density functions 
from spectropolarimetric measurements is not straight forward. 
There is no direct link to the number density of 
strong field components without accounting at the same time for the 
inherent length scale (correlation length) of the magnetic structures. 
The magnitude and the profile form of the polarized signal is therefore much more 
the result of the complex interplay between the horizontal probability density function 
and the characteristic length scale of the individual structures.

\section{Summary and Conclusions}
\label{Summary}
The ubiquity as well as the degree of asymmetry of observed Stokes V profiles 
are direct indicators of the inhomogeneous nature of the unresolved magnetic field
in the solar photosphere.
Stokes V profile asymmetries can be explained in terms of atmospheric gradients 
or more general in terms of the finite spatial extent (correlation length) 
of the underlying magnetic field.
This finite character of the individual structures is the essential starting point 
of the here presented concept of a mesostructured magnetic atmosphere (MESMA). 
In contrast to the proposed microstructured magnetic atmosphere 
\citep{SA96} the MESMA approach does not rely on the 
assumption of a solely microturbulent magnetic field, 
the MESMA concept rather allows all kinds of different characteristic 
length scales to be present in the atmosphere. 

We have put forward a very generic statistical model (Sect. \ref{Mesma}) 
which is based on a Markov random field. The key idea here was to
explicitly account for the spatial coherency of the underlying structures.
This model facilitates the derivation of a stochastic transport equation
for the conditional Stokes vector (Sect. \ref{StochTrans}).
By utilizing a modified evolution operator we could derive in analogy 
to the deterministic case a formal solution for the stochastic polarized transport 
equation (Sect. \ref{Formal}).
In Sect. \ref{MicroMacro} we have shown that the 
microstructured (MISMA) as well as the macrostructured approach are special 
limits of our more general mesostructured approach.
Hence, the here outlined concept of a mesostructured magnetic atmosphere presents 
the natural extension of the MISMA approach.
We could also show that asymmetric Stokes profiles are a natural 
consequence for an atmosphere where velocity and the magnetic fields 
are organized on finite scales. 
Moreover, our stochastic approach allowed us to determine
a direct relation between the characteristic length scale of the atmospheric 
structures and the degree of the asymmetry (Sect. \ref{Asymmetry}). 

One of the most important parameter in the MESMA approach is the 
correlation length (the mean length scale) of the underlying atmospheric 
structures. As the correlation length appears 
explicitly in the stochastic transfer equation for polarized light 
it can be used as a diagnostic parameter. 
In a number of numerical experiments we used exactly this parameter to 
demonstrate the importance of the correlation length for the polarized line formation.

In a first numerical experiment (Sect. \ref{TwoCompField}) we have analyzed the line
formation in a fluctuating atmosphere which is characterized by two kinds (ensembles) 
of different structures (magnetic and non-magnetic). Albeit this is a commonly
invoked scenario for the interpretation of unresolved magnetic structures 
our approach fully accounts for the lack of knowledge about the underlying 
organization and geometry. 
In these model calculations it could be demonstrated that 
the Stokes V profiles as well as the amplitude and area asymmetries 
have a clear functional dependence upon the correlation length of the 
underlying structures. 
We also showed (Sect. \ref{ReallyMicro}) that the 
effective microturbulent limit is already reached for a finite correlation length of 
approximately 15 km. On the other hand this means that for 
correlation lengths larger than 20 km the atmosphere can no longer be 
described in terms of a microstructured magnetic atmosphere.

An atmosphere which is characterized by an extended magnetopause
like structure was investigated in Sect. (\ref{MagPause}).
Again, our results demonstrate the distinct change of the
Stokes V profiles and other derived quantities with 
varying correlation length.
The line formation strongly depends on the structure below the 
magnetopause. The ratio of the Stokes V profile amplitude
and the ratio of the magnetic flux density for the
two iron line  Fe {\scriptsize I} $\lambda 6301 \AA \:$
and Fe {\scriptsize I} $\lambda 6302 \AA \:$ are significantly
altered just by changing the characteristic length scale
of the magnetic field. These ratios falsely indicate 
for small correlation lengths the presence of 
strong  kilo-Gauss magnetic fields, 
although the entire model atmosphere is intrinsically weak. 
Thus, the validity and application of these ratios 
as indicators for strong magnetic field structures 
is highly questionable in inhomogeneous atmospheres.
 
In Sect. (\ref{FieldDist}) we finally investigated the
polarized spectral signature which originated from a
field strength distribution given by a recently
derived empirical probability density functions.
We placed particular emphasis on the fact that the underlying 
magnetic field cannot properly be described in the context of 
a field strength distribution without taking into account the 
characteristic length scale of the individual structures. 
As the field strength covers a broad range from very weak 
to strong fields, the characteristic length scale of these
structures has to be different. If we realistically 
assume that strong field structures possess  
intrinsic length scales that are larger than that of the 
weaker field structures
we have shown that the strong field part of the distribution 
dominates in contributing to the resulting Stokes V profile and 
would lead to an overestimation of the strong field structures.
Therefore neglecting the correlation length (line-of-sight filling factor)
of the underlying structures will lead to incorrect estimates for the
magnetic field and flux distribution. 

The here introduced concept of a mesostructured magnetic atmosphere (MESMA) is a 
statistical attempt to cope with the underlying complexity of the magnetic field
in the solar photosphere in terms of polarized radiative transfer modeling.  
As the majority of the magnetic flux in the solar photosphere 
is still not resolved and the fundamental length scale 
of the magnetic structures is unknown, the here presented stochastic 
approach offers a viable and promising tool for the interpretation of 
spectropolarimetric observations and the diagnostics of photospheric 
magnetic fields.

First Stokes profile inversions on the basis of the MESMA approximation were made
by \citet{Car05b,Car06, Car07}. They have demonstrated the feasibility 
of their approach which allowed them to estimate the characteristic length scale
of internetwork and penumbral magnetic field structures.

\begin{acknowledgements}
We would like to thank the referee B. Ruiz Cobo for many helpful and constructive 
comments and suggestions that helped to improve this paper.
The authors also gratefully acknowledge financial support from the Deutsche
Forschungsgemeinschaft (DFG) under the grant CA 475/1-1. 
\end{acknowledgements}

\appendix
\section{The master equation for the mesostructured magnetic atmosphere}
\label{Appendix_Master}

In following we want to derive a differential equation (master equation) that
governs the spatial evolution of the probability density along 
the spatial coordinate $s$. 
To derive such a master equation for the 
Kubo-Andreson process let us take the Taylor expansion 
of the transition probability (\ref{kubo}) at the coordinate $s'$ 
which gives
\begin{eqnarray}
p(\bm{B}'',s'+\Delta s \mid \bm{B}',s') = \nonumber 
\left ( 1 - \gamma(\bm{B}') \Delta s \right )
\delta(\bm{B}'' - \bm{B}') +   \hspace{0.4cm} \nonumber \\
+ \; \gamma(\bm{B}') q(\bm{B}'') \Delta s' 
+ O(\Delta s'^2), 
\label{firstorder}
\end{eqnarray} 
where we have assumed that $\gamma(\bm{B}')$ is sufficiently small compared to 
$\Delta s$ such that second and higher order terms of the exponential
are negligible. 
We then introduce the transition rate  
$w(\bm{B}'',\bm{B}')$ which describes the probability per unit length scale 
for a transition from an atmospheric regime $\bm{B}'$ to a regime $\bm{B}''$. 
The transition rate can be obtained from the spatial derivative 
of the Kubo-Anderson Process (\ref{kubo}), hence, for a sufficiently small 
$\gamma(\bm{B})$ the transition rate directly follows from Eq. \ref{firstorder},
\begin{eqnarray}
w(\bm{B}'',\bm{B}') = \gamma(\bm{B}') q(\bm{B}'') \; . 
\label{transrate}
\end{eqnarray}
To introduce the dynamics of the process, we use the Chapman-Kolmogorov equation 
(\ref{chapkol}),
\begin{eqnarray}
p(\bm{B}'',s'+\Delta s \mid \bm{B},s ) \hspace{4.7cm} \nonumber \\
= \int p(\bm{B}'',s'+\Delta s \mid \bm{B}',s')
\; p(\bm{B}',s' \mid \bm{B},s) \; d\bm{B}' \; , 
\label{chapkol2}
\end{eqnarray}
and replace the first factor under the integral on the r.h.s. with 
Eq. (\ref{firstorder}) to obtain
\begin{eqnarray}
p(\bm{B}'',s'+\Delta s \mid \bm{B},s ) 
= \left (1 - \gamma(\bm{B}'') \Delta s \right ) 
p(\bm{B}'',s' \mid \bm{B},s) \hspace{0.7cm} \nonumber \\
+ \; \Delta s \int w(\bm{B}'',\bm{B}') p(\bm{B}',s' \mid \bm{B},s) \: d\bm{B}' 
\; ,\hspace{0.3cm}
\label{firstset}
\end{eqnarray}
Please note, that we have already performed the integration of the first term
over the delta function with respect to B'.  
We rearrange Eq. (\ref{firstset}) to write
\begin{eqnarray}
\frac{p(\bm{B}'',s'+\Delta s \mid \bm{B},s ) - 
p(\bm{B}'',s' \mid \bm{B},s)}{\Delta s} \hspace{1.8cm} \nonumber \\
=  - \gamma(\bm{B}'') p(\bm{B}'',s' \mid \bm{B},s) \hspace{2.3cm}  \nonumber \\
+ \; \int w(\bm{B}'',\bm{B}') p(\bm{B}',s' \mid \bm{B},s) \; d\bm{B}' 
\; . \hspace{-0.4cm} 
\label{diffquot1}
\end{eqnarray}
The definition of the transition rate (\ref{transrate}) allows us to write the 
fluctuation rate $\gamma(\bm{B}'')$ in the following form,
\begin{eqnarray}
\gamma(\bm{B}'') = \int w(\bm{B}',\bm{B}'') \: d\bm{B}' \; ,
\end{eqnarray}
note here, that $q$ in (\ref{transrate}) satisfy the normalization condition (\ref{normcond}).
This relation can now be used to write (\ref{diffquot1}) as 
\begin{eqnarray}
\frac{p(\bm{B}'',s'+\Delta s \mid \bm{B},s ) - 
p(\bm{B}'',s' \mid \bm{B},s)}{\Delta s} \hspace{1.8cm} \nonumber \\
 = \;  \int w(\bm{B}'',\bm{B}') p(\bm{B}',s' \mid \bm{B},s) \; d\bm{B}' \hspace{1.3cm} \nonumber \\
-  \int w(\bm{B}',\bm{B}'') p(\bm{B}'',s' \mid \bm{B},s) \; d\bm{B}' \; . \hspace{-0.3cm} 
\label{diffquot2}
\end{eqnarray}
Taking the limit as $\Delta s \rightarrow 0$ gives the differential equation which
describes the spatial evolution of the conditional probability density,
\begin{eqnarray}
\frac{\partial p(\bm{B}'',s' \mid \bm{B},s)}{\partial s'}  \hspace{4.8cm} \nonumber \\
 = \int w(\bm{B}'',\bm{B}') p(\bm{B}',s' \mid \bm{B},s) \; d\bm{B}' \hspace{1.0cm} \nonumber \\
- \int w(\bm{B}',\bm{B}'') p(\bm{B}'',s' \mid \bm{B},s) \; d\bm{B}'
\; . \hspace{-0.6cm} 
\label{cond_master}
\end{eqnarray}
To obtain the differential equation for the unconditional probability density function 
we multiply (\ref{cond_master}) with $p(\bm{B},s)$ followed by an integration 
over the entire state space $\bm{B}$. After rearranging the indices we 
finally yield the master equation for the Kubo-Anderson process,
\begin{eqnarray}
\frac{\partial p(\bm{B},s)}{\partial s}  = \int w(\bm{B},\bm{B}') 
p(\bm{B}',s') \; d\bm{B}' \hspace{2.2cm} \nonumber \\
\hspace{1.0cm} - \int w(\bm{B}',\bm{B}) p(\bm{B},s) \; d\bm{B}' \; .
\label{app_master}
\end{eqnarray}
The discrete analogue of the master equation is given by
\begin{eqnarray}
\frac{\partial p(\bm{B},s)}{\partial s}  = \sum_{B'} w(\bm{B},\bm{B}') 
p(\bm{B}',s') \hspace{2.2cm} \nonumber \\
\hspace{1.0cm} - \sum_{B'} w(\bm{B}',\bm{B}) p(\bm{B},s) \; ,  
\label{app_disc_master}
\end{eqnarray}
where the sum spans over the entire discrete state space of $\bm{B}'$.

\section{The wavelength integrated mean line depression contribution function 
for circular polarized light}
\label{Appendix_CFRCP}

The wavelength integrated mean line depression contribution function 
can be derived from the formal solution of the mean conditional Stokes vector (\ref{formal}) 
which reads for a semi-infinite atmosphere
\begin{equation}
\bm{Y_B}(0) \: = \: \int_0^{\infty} \bm{\tilde{O}_B}(0,\tau) \: \bm{\tilde{j}_B}(\tau)  d\tau \: ,
\label{formalapp}
\end{equation}
where $\bm{\tilde{j}_B}(\tau)$ is the modified conditional line emission vector (\ref{modemission}).
From the formal solution we can define the contribution function to the emergent mean conditional 
Stokes vector as 
\begin{eqnarray}
\bm{\tilde{C}_B}(\log \tau) \: = \: \ln(10) \: \tau \: \bm{\tilde{O}_B}(0,\tau) \: \bm{\tilde{j}_B}(\tau) \: .
\hspace{0.25cm}
\end{eqnarray}
To derive the line depression contribution function of the conditional Stokes vector we 
introduce the line depression of the conditional Stokes vector as
$\bm{R} = \bm{1} - \frac{\bm{Y}}{<I_c>}$. 
The derivation of the line depression contribution function is then formally identical to the non-stochastic 
case \citep[see e.g.][]{Sol94,Rees96}. 
By using again the formal solution (\ref{formalapp}) we can write for the line depression 
contribution function of the mean conditional Stokes vector
\begin{equation}
\bm{\tilde{C}_{BR}}(\log \tau) = \: \ln(10) \: \: \tau \: \: \bm{\tilde{O}_B}(0,\tau) \: \: \frac{ <I_c(\tau)> \: -
\:\bm{\tilde{j}_B}(\tau)}{<I_c(0)>} \: .
\end{equation}
The wavelength integrated mean line depression contribution function of the circular polarization can 
then finally be defined as
\begin{equation}
\tilde{C}_{R;CP}(\log \tau) \: = \: \int_{\Lambda} \int_{\hat{\bm{B}}}\mid \tilde{C}_{BR;V}(\bm{B},\lambda,\tau) 
\mid  p(\bm{B})\: d\bm{B} \: d\lambda \: ,
\label{CFRCP}
\end{equation}
where the integration of the inner integral is performed over the state space $\bm{\hat{B}}$
and the outer integration over the wavelength domain $\Lambda$ of the spectral line.

\end{document}